\documentclass[aip,amsmath,amssymb,reprint]{revtex4-1}
\draft
\usepackage{graphicx}  
\usepackage{dcolumn}
\usepackage{amsmath}
\usepackage{makecell}
\usepackage{bm}        
\usepackage{amssymb}   

\usepackage[utf8]{inputenc}
\usepackage[T1]{fontenc}
\usepackage{mathptmx}
\usepackage{ragged2e} 
\usepackage{float}
\usepackage{color}

\begin{document}

\title{Determining fluid-crystal phase boundaries for a binary hard-sphere mixture using direct-coexistence simulations}

\author{Rinske M. Alkemade}\email[Author to whom correspondence should be addressed: ]{r.m.alkemade@uu.nl}
\affiliation{Soft Condensed Matter and Biophysics, Debye Institute for Nanomaterials Science, Utrecht University, Utrecht, Netherlands }
\author{Alessandro Salo}
\affiliation{Soft Condensed Matter and Biophysics, Debye Institute for Nanomaterials Science, Utrecht University, Utrecht, Netherlands }
\author{Laura Filion}
\affiliation{Soft Condensed Matter and Biophysics, Debye Institute for Nanomaterials Science, Utrecht University, Utrecht, Netherlands }
\author{Frank Smallenburg}
\affiliation{
Universit\'e Paris-Saclay, CNRS, Laboratoire de Physique des Solides, 91405 Orsay, France
}

\begin{abstract}
Determining fluid–crystal phase boundaries via direct-coexistence methods can be challenging due to the fact that the simulation box can introduce crystal strain. Recently, a direct-coexistence approach was developed which allows one to easily identify the equilibrium strain-free fluid-crystal coexistence in monodisperse systems. Here, we show that this approach can be readily extended to binary mixtures forming stoichiometric binary crystals, allowing accurate and efficient determination of the phase boundaries. Moreover, we examine how the choice of crystal plane in contact with the fluid affects the accuracy of the phase boundary determination. The method is easy to implement and does not require prior knowledge of the binary fluid’s equation of state. These results further establish the method as a robust and practical tool for accurately determining fluid–crystal phase boundaries.
\end{abstract}

\maketitle

\section{Introduction}
When studying the behaviour of colloidal and atomic substances, it is essential to understand a system's phase diagram and its associated phase boundaries. Identifying the conditions under which phase transitions occur is a crucial first step toward understanding a range of related physical processes, such as nucleation and interfacial behaviour.

Computer simulations are a powerful tool for determining phase boundaries\cite{vega2008determination, frenkel2023understanding, chew2023phase}. Several methods exist to determine the thermodynamic conditions under which two or more phases can coexist, meaning the phases have the same pressure $P$, chemical potential $\mu$ and temperature $T$. These methods include brute-force approaches, which directly explore which phases occur at specific state points, and free-energy-based methods\cite{frenkel2023understanding, frenkel1984new,bolhuis1997tracing, schilling2009computing, polson2000finite, vega2007revisiting, dijkstra2014entropy}, which compute the bulk free energy of each phase to determine coexistence conditions. In this paper, we focus on a third approach that considers direct-coexistence simulations\cite{opitz1974molecular,ladd1977triple,ladd1978interfacial,cape1978molecular}. In this class of methods, both coexisting phases are simulated within a single system such that the interface between the phases is incorporated explicitly. Since particles, energy, and volume can redistribute between the phases, the coexistence criteria are automatically satisfied, giving direct access to the equilibrium properties of each phase.

However, identifying fluid-crystal coexistences via direct-coexistence simulations can be challenging because crystal strain can be induced by the shape of the simulation box\cite{broughton1986molecular, noya2008determination, espinosa2013fluid}. To address this, Ref.~\onlinecite{smallenburg2024simple} recently proposed a straightforward approach to determine the equilibrium fluid-crystal coexistences. In order to identify the unstrained crystal, they perform a series of direct-coexistence simulations across crystal densities. The unstrained equilibrium crystal is then identified as the one whose associated bulk pressure matches the pressure as measured normal to the interface in the direct-coexistence simulation.

The goal of this paper is to extend the approach of Ref.~\onlinecite{smallenburg2024simple}, which was tested on a collection of monodisperse systems, to stoichiometric binary crystals. In comparison to monodisperse systems, identifying the coexisting phases in binary systems presents an additional challenge because the composition can differ between phases. 
If the composition of both phases is unknown a priori, determining the phase boundaries via direct-coexistence methods requires extensive sampling over many possible particle mixtures or the use of other more advanced techniques\cite{castagnede2025freezing}. In stoichiometric crystals, however, the crystal composition is fixed by the crystal structure (assuming a low defect concentration). During a direct-coexistence simulation, the fluid can freely exchange particles with the crystal at the interface, allowing the fluid composition to adjust until both phases reach equilibrium. As a result, the simulation naturally converges to the correct coexistence composition, implying that the method of Ref. \onlinecite{smallenburg2024simple} can be readily applied to stoichiometric binary crystals.

Here, we test this approach by reconstructing the phase diagram of a binary hard-sphere mixture with size ratio $q = 0.58$, which has previously been predicted via free-energy calculations by Eldridge, Madden, and Frenkel \cite{eldridge1993entropy}. We demonstrate that the method is simple to implement and yields highly accurate phase boundaries. Additionally, we comment on the effects of the choice of crystal plane in contact with the fluid on the accuracy of this method.

\begin{figure*}
  \begin{minipage}[t]{0.21\linewidth}
        \RaggedRight a)\\
        \includegraphics[height=3.5cm]{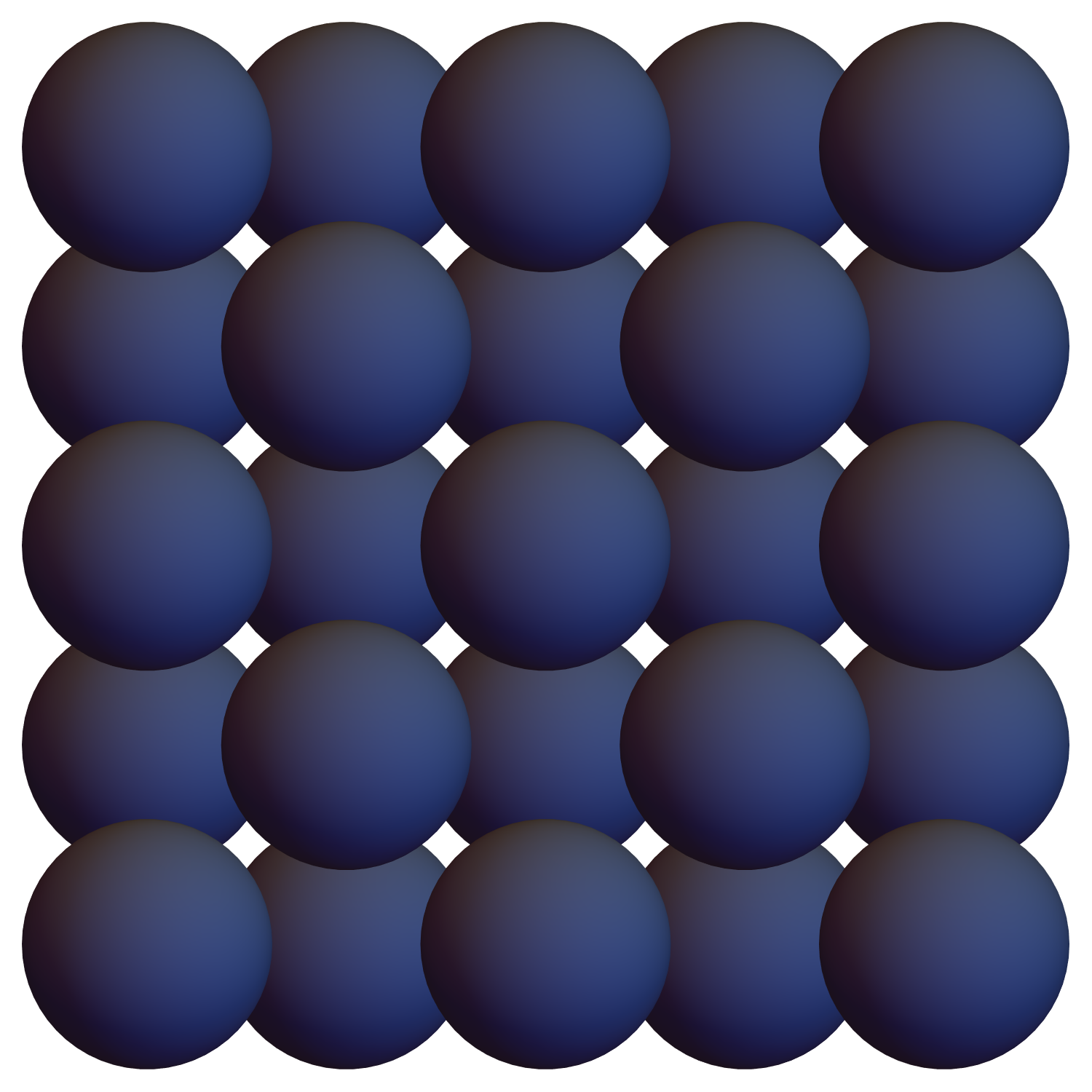}
    \end{minipage}
    \begin{minipage}[t]{0.23\linewidth}
        \RaggedRight b)\\
        \includegraphics[height=3.5cm]{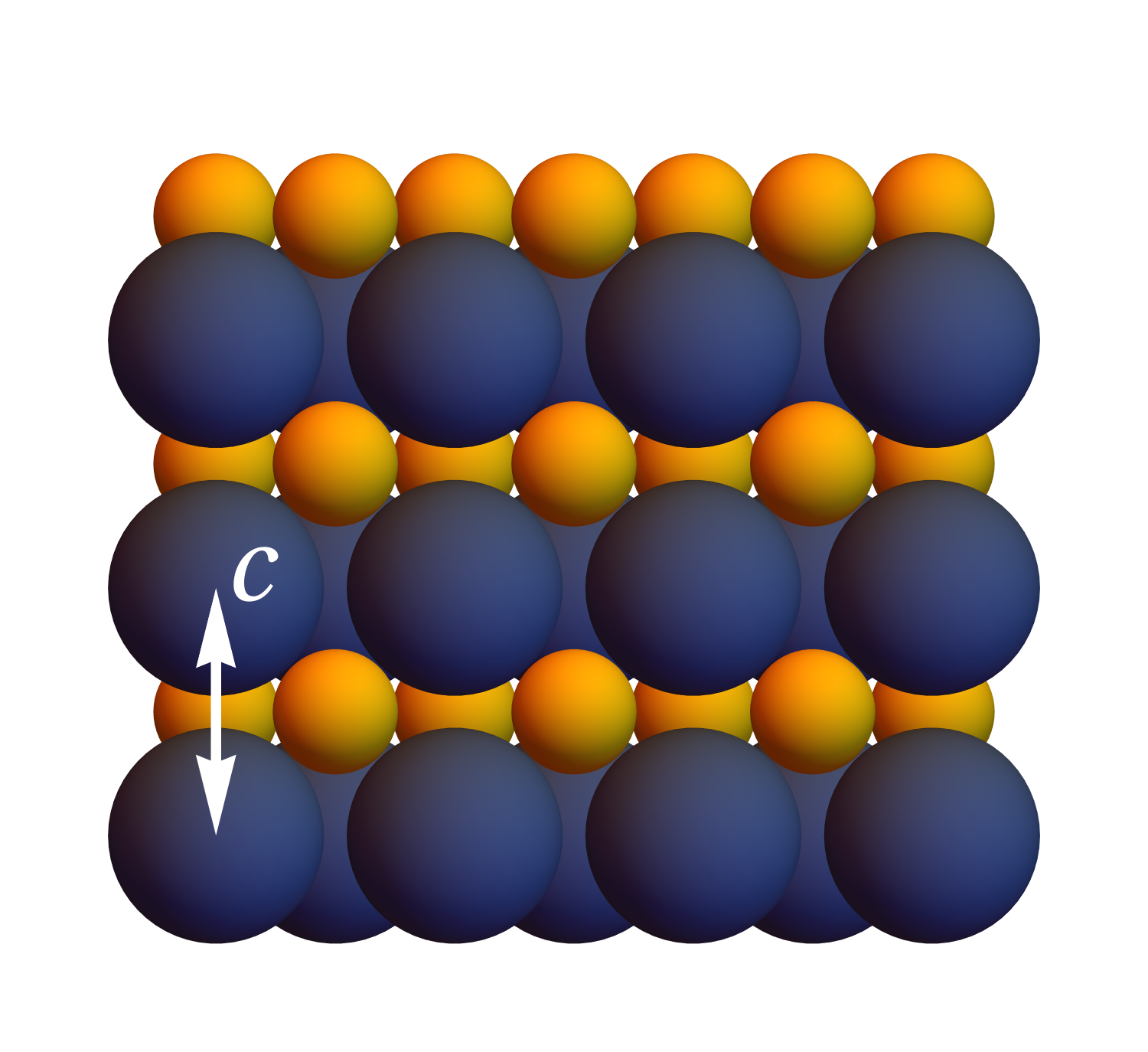}
    \end{minipage}
     \begin{minipage}[t]{0.23\linewidth}
     \RaggedRight \text{ }\\
        \includegraphics[height=3.5cm]{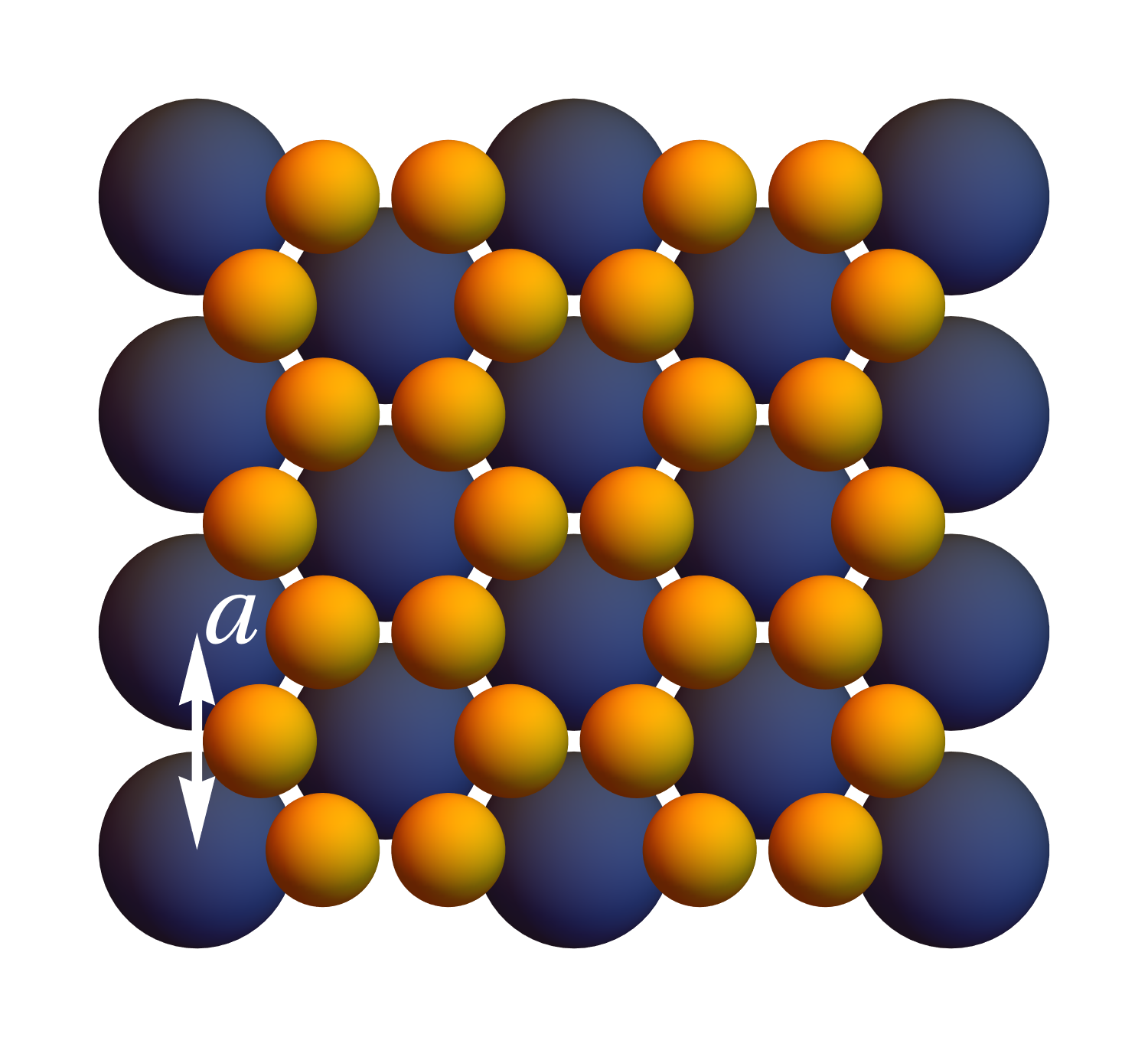}
    \end{minipage}
   \begin{minipage}[t]{0.29\linewidth}
        \RaggedRight c)\\
        \includegraphics[height=3.5cm]{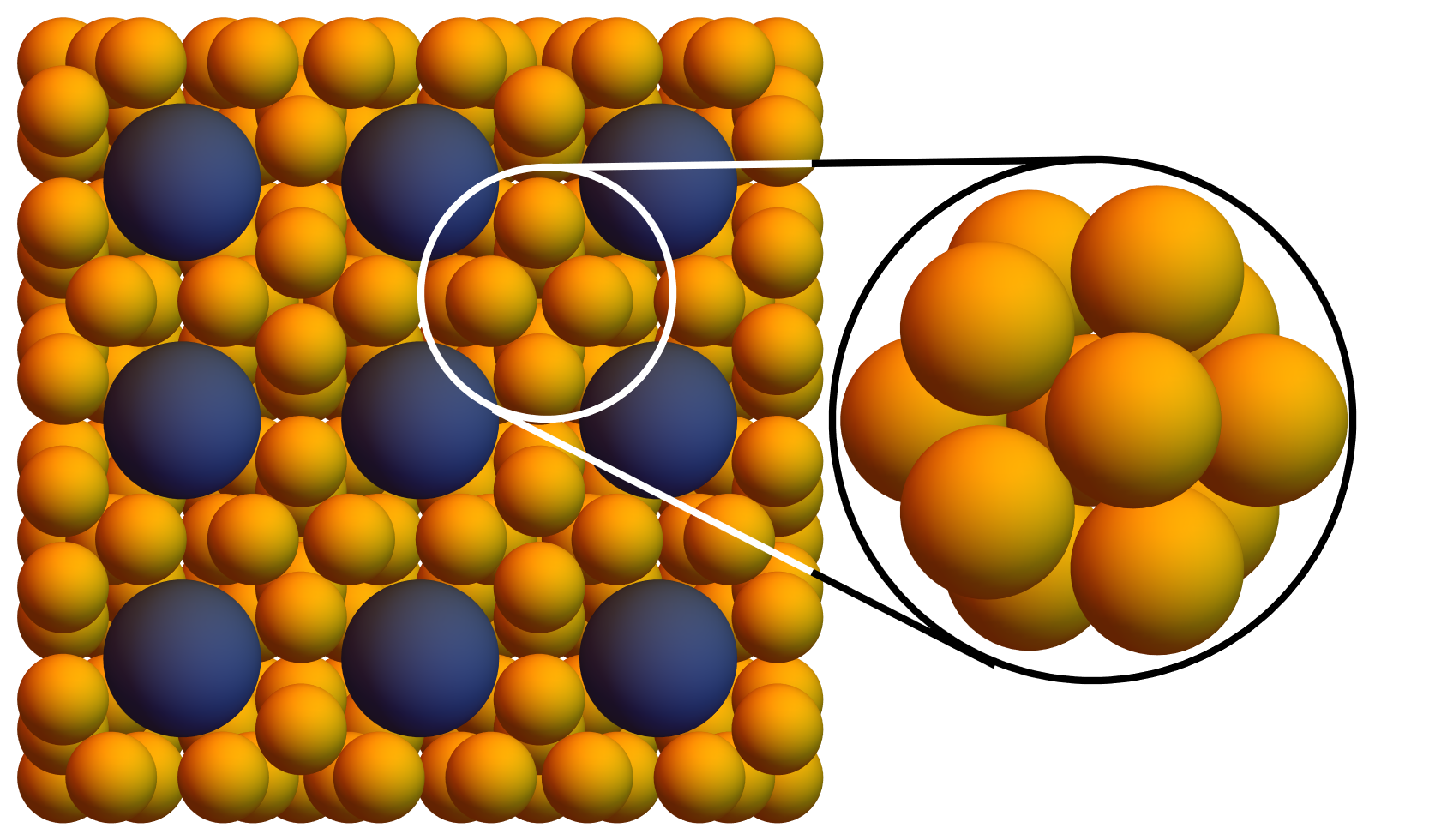}
    \end{minipage}
    
    \caption[width=1\linewidth]{Crystals that can form in a hard-sphere binary mixture at size ratio $q=0.58$. a) An FCC(A) crystal consisting of $A$-particles. Note that the structure of the FCC(B) crystal (not displayed here) is identical to the FCC(A) crystal.  b) An AB$_2$ crystal, consisting of alternating layers of $A$ and $B$ particles, denoted by respectively the blue and yellow particles. The large particles form hexagonal layers and the small particles form a planar honeycomb lattice. We denote the lattice spacing associated with the hexagonal plane by $a$, while $c$ denotes the lattice spacing between consecutive layers of the same type. c) An AB$_{13}$ crystal consisting of a cubic lattice of $A$-particles, with each cubic cell containing an isocahedral cluster of 13 $B$-particles. Neighbouring isocahedral structures are rotated by $\pi/2$, such that the total unit cell of AB$_{13}$ consists of 112 particles.}
    \label{fig:crystals} 
\end{figure*}

\section{Methods}

\subsection{Binary hard-sphere mixture}
As our model system, we use a binary hard-sphere mixture consisting of particle types $A$ and $B$. These particles have a size ratio $q = \sigma_B/\sigma_A =0.58$, where $\sigma_i$ denotes the diameter of species $i$. Note that these hard spheres interact solely via excluded-volume interactions, i.e. they cannot overlap, but otherwise they do not interact. The binary mixture under consideration has been extensively studied \cite{eldridge1993entropy,eldridge1995binary} and is known to form four different crystal phases that can all coexist with a binary fluid. Fig. \ref{fig:crystals} shows the observed crystals, which consist of monodisperse  FCC(A/B) crystals of pure $A$ and $B$ particles and binary crystals, AB$_2$  (atomic analogue AlB$_2$)  and AB$_{13}$ (atomic analogue NaZn$_{13}$) .

We simulate the system using event-driven molecular dynamics simulations\cite{alder1960studies, rapaport2004art} (EDMD) implemented in the canonical ensemble, where the temperature is fixed using an Andersen thermostat\cite{frenkel2023understanding}. The EDMD code we use is adapted from Ref.~\onlinecite{smallenburg2022efficient} to allow for measurements of pressure tensors and composition profiles. 
The simulation consists of $N = N_A + N_B$ particles placed in a box with periodic boundaries, where $N_i$ is the number of particles of species $i$. Here, we choose to express the composition in terms of the fraction of small particles, i.e. the global composition is equal to $\chi^G = N_B/(N_A + N_B)$. During the simulation, the pressure tensor $P_{kl}$ is measured by tracking the momentum transfer at each collision~\cite{alder1960studies},
\begin{equation*}
    \frac{\beta P^G_{kl}}{\rho} = \delta_{kl} - \frac{\sum_{c}m\,\delta v^{(k)}_{i} r^{(l)}_{ij}}{N\,\Delta t} ,
\end{equation*}
where $\beta=1/k_BT$ with $k_B$ Boltzmann's constant, $\rho = N/V$ is the number density, and the sum runs over all collisions, $c$, occurring within the time window $\Delta t$. At each collision, $\mathbf{r}_{ij} = \mathbf{r}_j - \mathbf{r}_i$ is the vector connecting the two colliding particles $i$ and $j$, and $m \delta \mathbf{v}_i$ is the momentum change of particle $i$, with $m$ the particle mass (which we choose identical for all particles). Additionally, the superscript $^{(s)}$ denotes the $s$-component of that vector, with $s\in\{x,y,z\}$.

\subsection{Direct coexistence}
We briefly summarize the method introduced in Ref.~\onlinecite{smallenburg2024simple} for determining fluid-crystal phase boundaries using direct-coexistence simulations. In such simulations, the coexisting phases are placed in a simulation box that is elongated along one direction (here chosen to be the $z$-direction). Due to free-energy minimization, the fluid-crystal interfaces will minimize and thus align perpendicular to the $z$-axis. As a result, mechanical equilibrium ensures that the normal pressure in that direction, $P^G_{zz}$, is unaffected by the interfaces and thus is the same inside both phases $\alpha$ and $\beta$, i.e. 
\begin{equation*}
P^G_{zz}= P^\alpha_{zz}= P^\beta_{zz},    
\end{equation*} 
with $P^\alpha_{zz}$ and $P^\beta_{zz}$ the local pressure in phase $\alpha$ and $\beta$ respectively. 

As mentioned, a key challenge in fluid–crystal direct-coexistence simulations is that the simulation box can impose strain on the crystal. This strain can arise because periodic boundaries fix the initial crystal lattice spacing in the $x$- and $y$-directions, which in turn fixes the initial crystal density $\rho^X_0$. If this $\rho^X_0$ does not correspond to the density of the crystal at the equilibrium melting point, the coexisting crystal will undergo a uniaxial strain of $\epsilon_{zz}$ along the $z$-axis. Since the pressure along the $z$-axis must be spatially uniform, the global pressure in that direction can thus be written as
\begin{equation*}
    P_{zz}^G(\rho_0^X, \epsilon_{zz}) = P^{UD}(\rho^X_0) + B_{zzzz}(\rho^X_0)\cdot\epsilon_{zz},
\end{equation*}
with $P^{UD}(\rho^X_0)$ the pressure associated with an undeformed crystal at density $\rho^X_0$, and $B_{zzzz}(\rho^X_0)$ the effective elastic constant of that same crystal, associated with a compression/expansion along the $z$-axis \cite{ray1989effective}. 

Ref. \onlinecite{smallenburg2024simple} showed that the unstrained coexistence, which corresponds to the situation where $\epsilon_{zz}=0$, can easily be identified by finding the density $\rho_0$ for which
\begin{equation*}
    P_{zz}^G(\rho_0^X) = P^{UD}(\rho^X_0). 
\end{equation*}
In practice, the equilibrium density $\rho_0^{X, UD}$  is determined by performing a short series of independent direct-coexistence simulations at varying $\rho_0^X$ and identifying the density for which the measured global pressure in the $z$-direction matches the undeformed bulk crystal pressure at the same density.
 
To apply this method to binary systems, we need to additionally consider the composition of the two coexisting phases. Here, we focus on coexistences involving stoichiometric crystals, for which the composition is known a priori. In other words, we assume a negligible crystal defect concentration, such that the crystal composition, $\chi^X$, remains fixed at its initial value. Note that this also implies that the pressure of the undeformed crystal can simply be determined from simulations of defect-free crystals. In contrast, the fluid composition $\chi^F$ can vary due to crystal melting or growth during the direct-coexistence simulation. Therefore, we measure the average $\chi^F(\rho_0^{X})$, and identify the equilibrium composition as that of the fluid coexisting with the undeformed crystal $\chi^F(\rho_0^{X, UD})$. 

\subsection{Composition measurements }
To measure the average $\chi^F(\rho_0^{X})$, we monitor the composition profile, $\chi(z)$, along the elongated $z$-axis during the direct-coexistence simulations. To avoid artifacts in the crystal composition, we partition the simulation box into discrete bins along the $z$-axis, where each bin has a width equal to the length of crystal unit cell in the $z$-direction. As the box length in $z$ generally does not accommodate an integer number of unit cells, we discard the final (incomplete) bin to maintain consistency.

Since the fluid-crystal interface can shift during the simulation, we align the composition profiles before averaging. To do so, we identify the middle of the high-composition phase and roughly align its position to $z/\sigma_A=0$. To compute the required shift necessary to align a profile, we perform a Fourier transform of the profile $\chi(z)$ using a wavelength equal to the box length in the $z$-direction, $L_z$. Specifically, we compute:
\begin{equation*}
\chi_k = \sum_i^{N_b} (\chi(z_i)-\chi_X)e^{-2\pi i \frac{z_i}{L_z}},
\end{equation*}
where $N_b$ is the number of integer bins in the box, $z_i$ is the $z$-coordinate of bin $i$, $\chi(z_i)$ is the local composition, and $\chi_X$ is the composition of the crystal phase. Given the resulting Fourier transform, we compute the associated phase $\phi = \arg(\chi_k)$, and shift each profile by an integer number of bins given by $\left\lfloor \frac{\phi L_z}{2\pi \cdot \text{binsize}} \right\rfloor$. The average composition profile, $\langle \chi(z) \rangle$, is then obtained by averaging over all the aligned profiles.

To identify the fluid region of the composition profile, we first split the profile into two halves, each containing one fluid–crystal interface. To each interface, a hyperbolic tangent function $\tanh(z)$ is fitted. The fluid is then defined as the region where the fitted function reaches at least $99.5\%$ of its maximum value (if $\chi^F > \chi^X$) or minimum value (if $\chi^F < \chi^X$). Finally, the fluid composition is determined by averaging over all the composition values within this fluid region.

\begin{figure}
        \includegraphics[width=\linewidth]{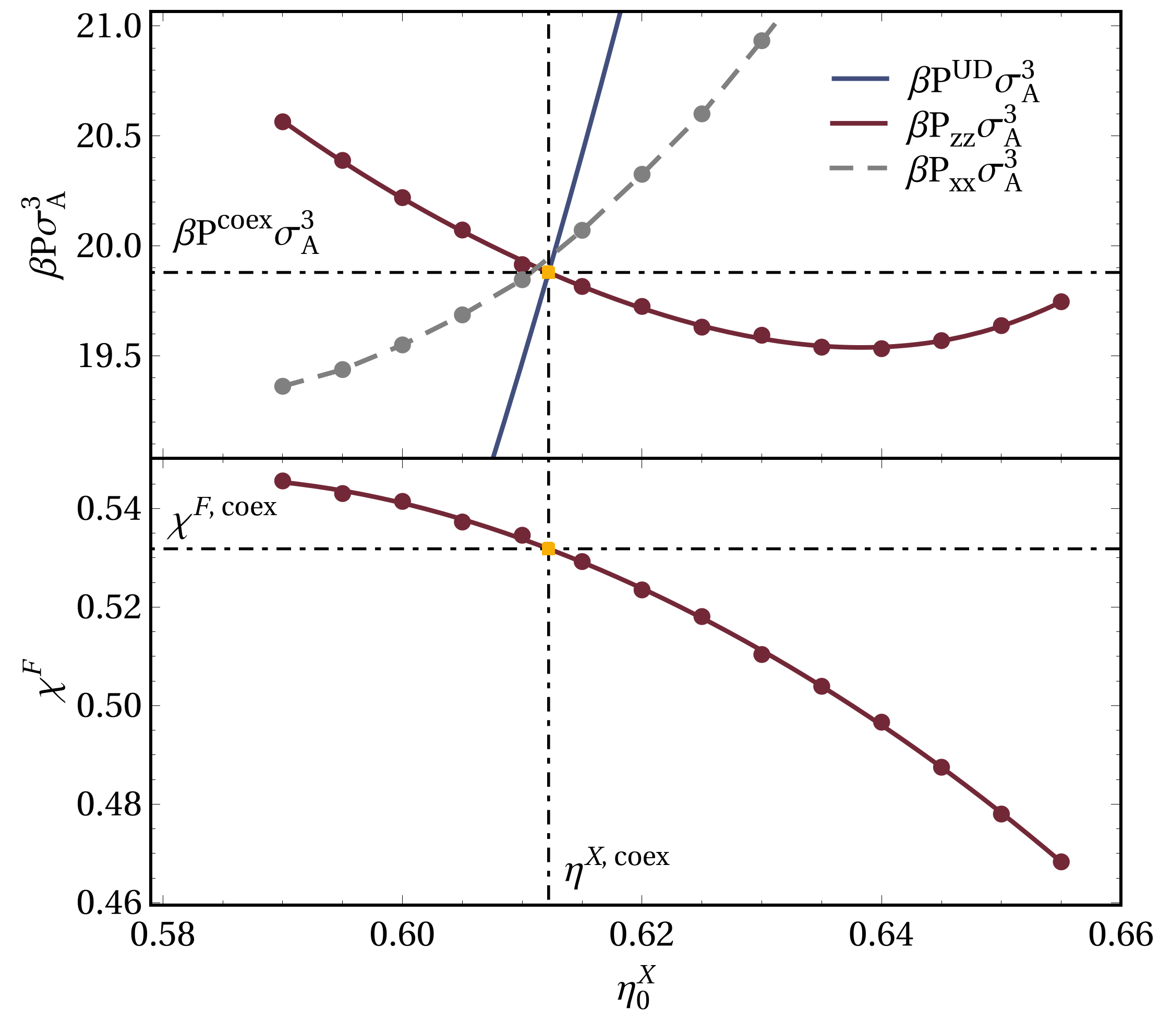}
    \caption[width=1\linewidth]{Demonstration of the direct-coexistence approach for a coexistence between a binary fluid and an FCC(A) crystal for a system with global packing fraction $\eta^G = 0.56$ and global composition $\chi^G = 0.3$. The upper panel shows the average pressure, $P_{zz}$, normal to the interface as a function of the initial crystal packing fraction $\eta^X_0$ (red line). In that same plot we plot the bulk equation of state $P^{UD}$ (blue line), measured in separate simulations, and the pressure $P_{xx}$ parallel to the interface of the direct-coexistence simulation (dashed grey line). In the plot, $P_{yy}$ is not shown, as it fully overlaps with $P_{xx}$. The coexistence pressure and crystal packing fraction are determined by identifying the crossing between $P^{UD}$ and $P_{zz}$ (yellow dot). In the lower panel, we show the fluid composition $\chi^F$ as a function of the initial crystal packing fraction. The coexistence fluid composition is equal to  $\chi^F(\eta^{X, \text{coex}})$.}
    \label{fig:intersectionFCC} 
\end{figure}

\begin{figure*}
  \begin{minipage}[t]{0.45\linewidth}
        \RaggedRight a)\\
        \includegraphics[width=\linewidth]{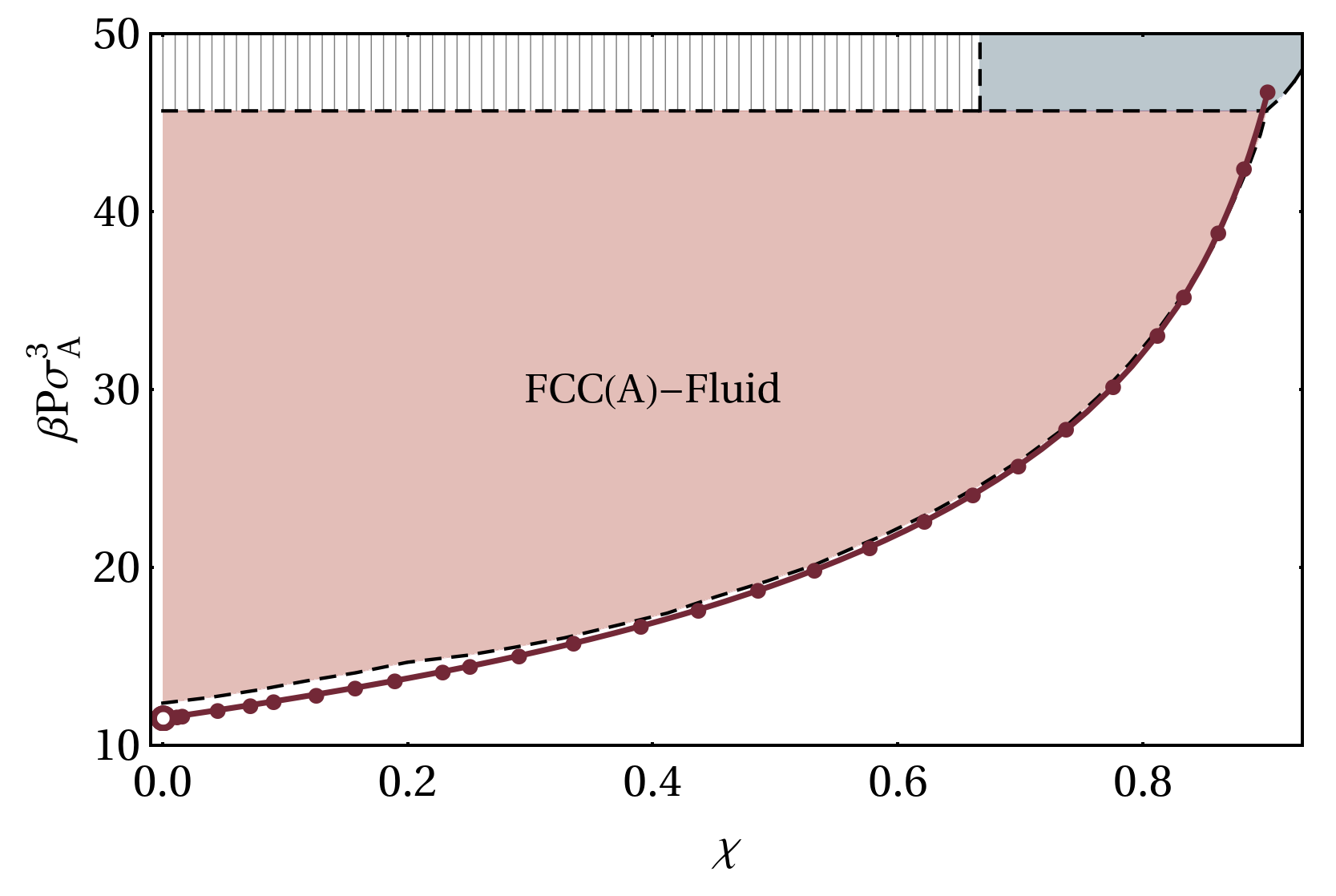}
    \end{minipage}
    \begin{minipage}[t]{0.45\linewidth}
        \RaggedRight b)\\
        \includegraphics[width=\linewidth]{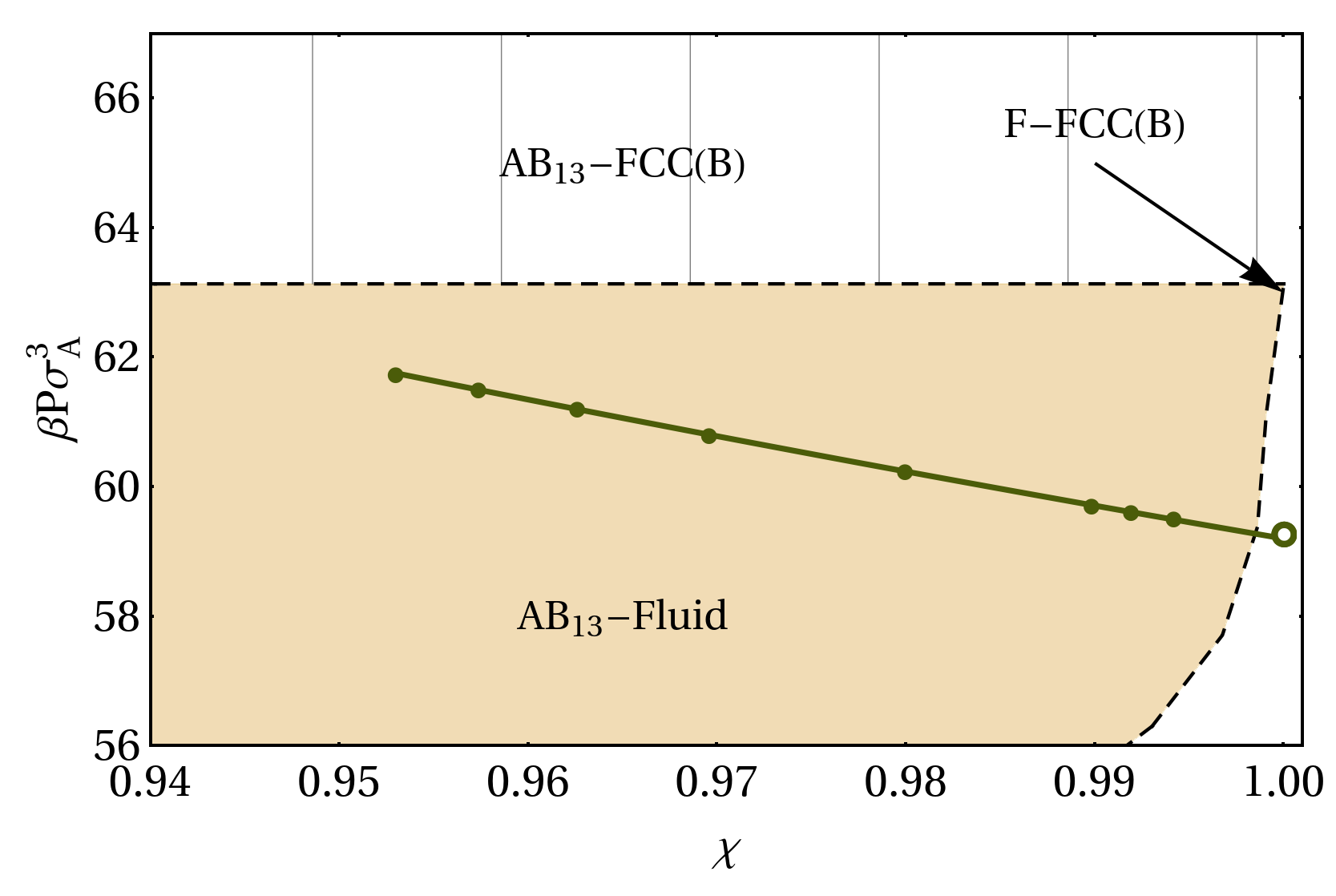}
    \end{minipage}
    \begin{minipage}[t]{0.45\linewidth}
        \RaggedRight c)\\
        \includegraphics[width=\linewidth]{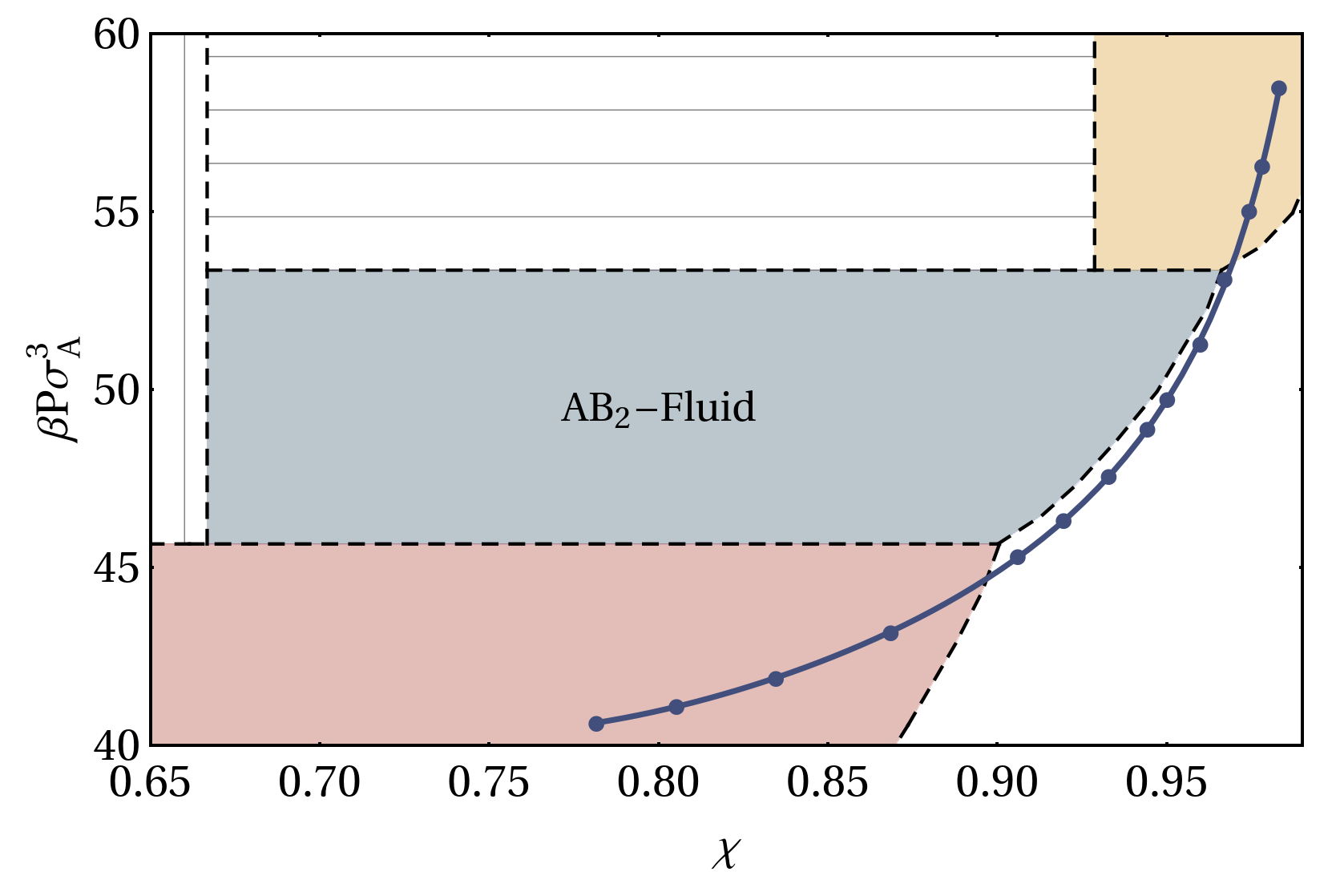}
    \end{minipage}
    \begin{minipage}[t]{0.45\linewidth}
        \RaggedRight d)\\
        \includegraphics[width=\linewidth]{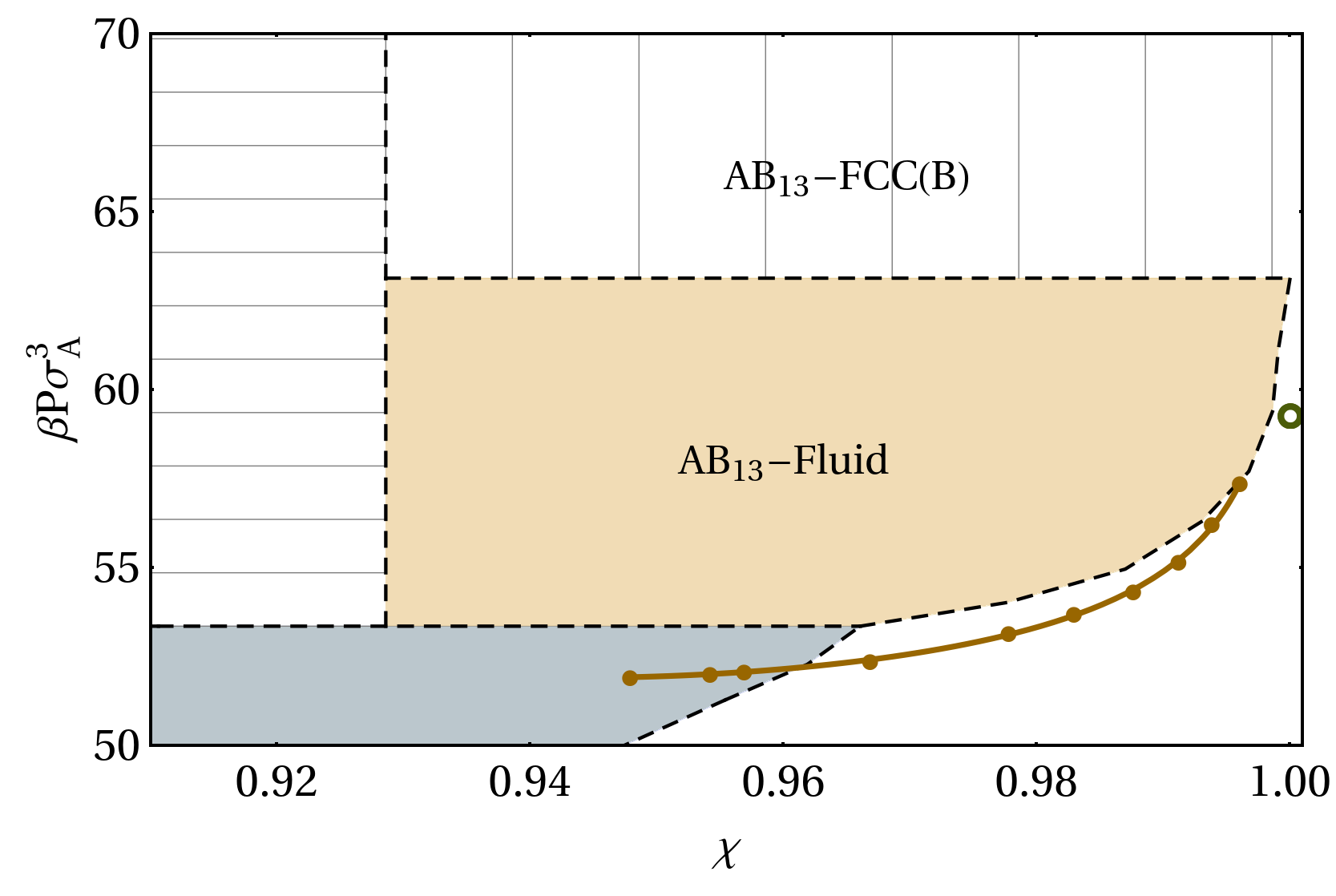}
    \end{minipage}
    \caption[width=1\linewidth]{Phase boundaries between a binary fluid and respectively: a) an FCC(A) crystal, b) an FCC(B) crystal, c) an AB$_2$ crystal, and  d) an AB$_{13}$ crystal. Plot markers represent results obtained using the direct-coexistence method, while dashed lines indicate the phase boundaries from Ref.~\onlinecite{eldridge1993entropy}. The solid lines are fitted functions through the direct-coexistence measurements (fits can be found in Appendix~\ref{app:fittedpolynomials}). The open markers denote the coexistence pressure between a monodisperse fluid and the FCC(A) or FCC(B) crystal, based on the results of Ref. \onlinecite{smallenburg2024simple}. Note that there is a large discrepancy between the location of the fluid–FCC(B) phase coexistence reported in Ref.~\onlinecite{eldridge1993entropy} and that obtained from the direct-coexistence simulations. Specifically, the phase boundaries in Ref.~\onlinecite{eldridge1993entropy} do not correctly terminate at the monodisperse fluid-crystal coexistence.}
    \label{fig:phasediagramperphase} 
\end{figure*}
\begin{figure}
  \begin{minipage}[t]{0.48\linewidth}
        \RaggedRight a)\\
        \includegraphics[width=\linewidth]{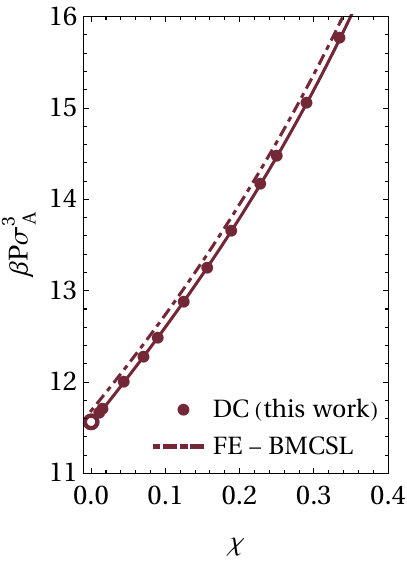}
    \end{minipage}
    \begin{minipage}[t]{0.48\linewidth}
        \RaggedRight b)\\
        \includegraphics[width=\linewidth]{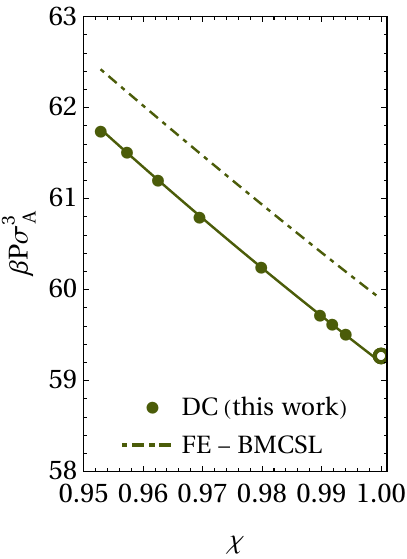}
    \end{minipage}
   
    \caption[width=1\linewidth]{Phase boundaries for a) fluid-FCC(A) and b) fluid-FCC(B), obtained from free-energy calculations that use the BMCSL EOS for the binary fluid (FE - BMCSL) and direct-coexistence simulations (DC). Note that the solid lines correspond to the solid lines plotted in Fig.~\ref{fig:phasediagramperphase}a) and b). Open markers denote the coexistence pressure between a monodisperse fluid and the FCC(A) or FCC(B) crystal, based on the results of Ref. \onlinecite{smallenburg2024simple}}
    \label{fig:freenergylines} 
\end{figure}

\subsection{Simulation details}
To redraw the phase diagram, we first determined the bulk equation of state, $P^{UD}(\rho^X)$ for all four crystals, using system sizes between $N=3000$ and $N=4000$, where $N$ is the number of particles in the system. Each simulation ran for $10^5\tau$, with $\tau = \sqrt{\beta m\sigma_A^2}$. Special care was required for the AB$_2$ crystal, since its unit cell is defined by two lattice parameters $a$ and $c$, see Fig. \ref{fig:crystals}. To determine the equilibrium $c/a$ ratio, we performed a series of bulk crystal simulations at various $c/a$ ratios for each crystal density, measuring the pressure tensor in each simulation. For each density, we then identified the $c/a$ ratio at which the pressure tensor becomes isotropic. In Appendix~\ref{app:latticeratio}, we show the ratio $c/a$, corresponding to the strain-free crystal, as a function of packing fraction $\eta$. The resulting fit for $c/a$ as a function of $\eta$ was used to set the lattice parameters in all AB$_2$ simulations. 

To initialize the direct-coexistence simulations, we selected, by trial and error, global packing fractions $\eta^G$ and global compositions $\chi^G$ that lied within a coexistence region, and that yielded coexistences with approximately equal crystal and fluid volumes. Note that for the system studied here, this procedure was relatively straightforward since the phase diagram was known beforehand. In the absence of prior knowledge of the phase diagram, it might be necessary to first perform exploratory bulk simulations over a range of densities and compositions in order to identify regions where phase separation could occur. For each state point that we considered, we picked a range of initial crystal packing fractions $\eta_0^X$, where for each value, we selected a corresponding fluid packing fraction $\eta^F$ and composition $\chi^F$ such that the desired $\eta^G$ and $\chi^G$ were achieved. The number of crystal particles, $N_X$, was set between $3500$ and $5500$, depending on the specific crystal. Given this $\eta_0^X$ and $N_X$, we computed the crystal volume $V_X$ and then chose the number of fluid particles such that the initial fluid volume $V_F$ matched $V_X$. To avoid initial overlap between the fluid and crystal, the fluid was first initialized in a box that was slightly smaller than $V_F$. After initialization, the fluid and crystal were glued together along the $z$-axis to form the initial direct-coexistence configuration.

We ran the direct-coexistence simulations for total times ranging from $10^6 \tau$ to $8\cdot10^6 \tau$, of which one-fourth was used as equilibration time. After equilibration, we started the measurement of the pressure tensor, and periodically measured the composition profile at time intervals at $\Delta t/\tau =1$. Shifting and averaging of the composition profiles were performed during the simulation.


\section{Results}
Here, we redraw the phase diagram for a binary hard-sphere mixture at size ratio $q=0.58$ in the pressure-composition representation. Results are compared to the phase diagram predicted by Eldridge, Madden, and Frenkel\cite{eldridge1993entropy}, where the data is extracted from their published phase diagram. Below, we discuss all four fluid–crystal coexistence branches separately.

\subsection{FCC(A) and FCC(B)}
To determine the fluid–FCC(A/B) phase boundaries, we performed direct-coexistence simulations across the relevant regions of parameter space. For the fluid-FCC(A) boundary this meant performing simulations for global packing fractions in the range $\eta^G\in[0.51, 0.61]$ and global compositions in the range $\chi^G\in[0.005, 0.6]$. To determine the fluid-FCC(B) phase boundary, we performed direct-coexistence simulations for global packing fractions in the range $\eta^G\in[0.52, 0.53]$ and global compositions in the range $\chi^G\in[0.975, 0.998]$. In both cases, the number of initial crystal particles was set to $N_X = 3840$, and the number of crystal unit cells in $x$- and $y$-direction was set to $8$. The FCC crystals were oriented with the (100) plane facing the fluid.

In Fig. \ref{fig:intersectionFCC}, we show how we determine the FCC(A) coexistence crystal packing fraction, $\eta^{X, \text{coex}}$, pressure, $P^\text{coex}$, and  composition, $\chi^{F, \text{coex}}$, using a series of simulations at global packing fraction $\eta^G = 0.56$ and global composition $\chi^G = 0.3$. It is noteworthy that, in contrast to the monodisperse system, the crossing between $P^{UD}(\eta_0^X)$ and $P^G_{zz}(\eta_0^X)$ is not associated with a minimum in $P^{UD}(\eta_0^X)$. In the monodisperse system, we see this crossing occur at a minimum because there is only one fluid and thus one equilibrium coexistence pressure, which corresponds to an unstrained crystal. Any strained crystal is by definition less stable, and hence must have a higher coexistence pressure\cite{smallenburg2024simple}. However, in a binary system, the crystal can coexist with a range of fluids, each at different pressures and compositions. Since varying $\eta_0^X$ changes the fluid composition, the equilibrium crystal is generally not associated with a minimum in pressure.

In Fig.~\ref{fig:phasediagramperphase}a) and b), we show the fluid–FCC(A/B) phase boundary obtained from direct-coexistence simulations respectively. For comparison, we include the phase boundaries from Ref.~\onlinecite{eldridge1993entropy}, obtained via free-energy calculations. In the figure, the monodisperse limits of the fluid–FCC(A/B) system\cite{smallenburg2024simple} --- which occur at $\beta P\sigma_A^3=11.5645$ and $\beta P\sigma_A^3=11.5645/q^3$,respectively  ---  are also included. Note, furthermore, that all binary coexistence points included in Fig.~\ref{fig:phasediagramperphase}b) correspond to metastable coexistences, as the region where fluid-FCC(B) coexistences are stable is extremely narrow. 

In both figures, we observe that the phase boundaries obtained from direct-coexistence simulations lie consistently below the phase boundaries predicted by Ref.~\onlinecite{eldridge1993entropy}. This discrepancy could be partly attributable to digitization errors introduced during data extraction. However, the systematic downward shift of phase boundaries is observed across (almost) all fluid-crystal pairs we studied. To investigate the origin of this shift, we recalculated the coexistence lines for FCC(A/B) using a reference Helmholtz free energy for the FCC crystal obtained by Polson, Trizac, Pronk, and Frenkel\cite{polson2000finite}, the FCC equation of state from Speedy \cite{speedy1998pressure}, and the semi-empirical fluid equation of state (EOS) by Boublik, Mansoori, Carnahan, Starling and Leland (BMCSL)\cite{boublik1970hard,mansoori1971equilibrium}. Note that the latter EOS was also used in Ref.~\onlinecite{eldridge1993entropy}. As shown in Fig.~\ref{fig:difPandF} in Appendix~\ref{app:discrepancyBMCSL}, using the BMCSL EOS leads to a systematic underestimation of both the pressure and Helmholtz free-energy of the fluid compared to simulations --- an observation that is consistent with earlier findings \cite{barovsova1996computer, santos1999equation, heyes2018chemical, castagnede2025freezing}. 
We plot the coexistence pressures recalculated using the BMCSL EOS, together with our direct-coexistence results in Fig. \ref{fig:freenergylines}. Consistent with our earlier observations, we see that the free-energy predictions consistently underestimate the coexistence pressure, even in the monodisperse limit. In this limit, it is easy to check the impact of the inaccuracies in the BMCSL EOS on the coexistence pressure by recalculating the coexistence again with a more accurate EOS for monodisperse hard spheres. Using the highly accurate Kolafa–Lab\'{i}k–Malijevsk\'{y} (KLM) EOS, we find excellent agreement with the direct-coexistence results. This indicates that the BMCSL EOS underestimates the monodisperse coexistence pressure by roughly 1\%, consistent with the overall shift shown in Fig. \ref{fig:freenergylines}.

\begin{figure*}

    \begin{minipage}[t]{0.49\linewidth}
        \RaggedRight a)
    \end{minipage}
    \begin{minipage}[t]{0.49\linewidth}
        \RaggedRight d)
    \end{minipage}
    
     \begin{minipage}[c]{0.49\linewidth}
     \centering
     \includegraphics[width=0.6\linewidth]{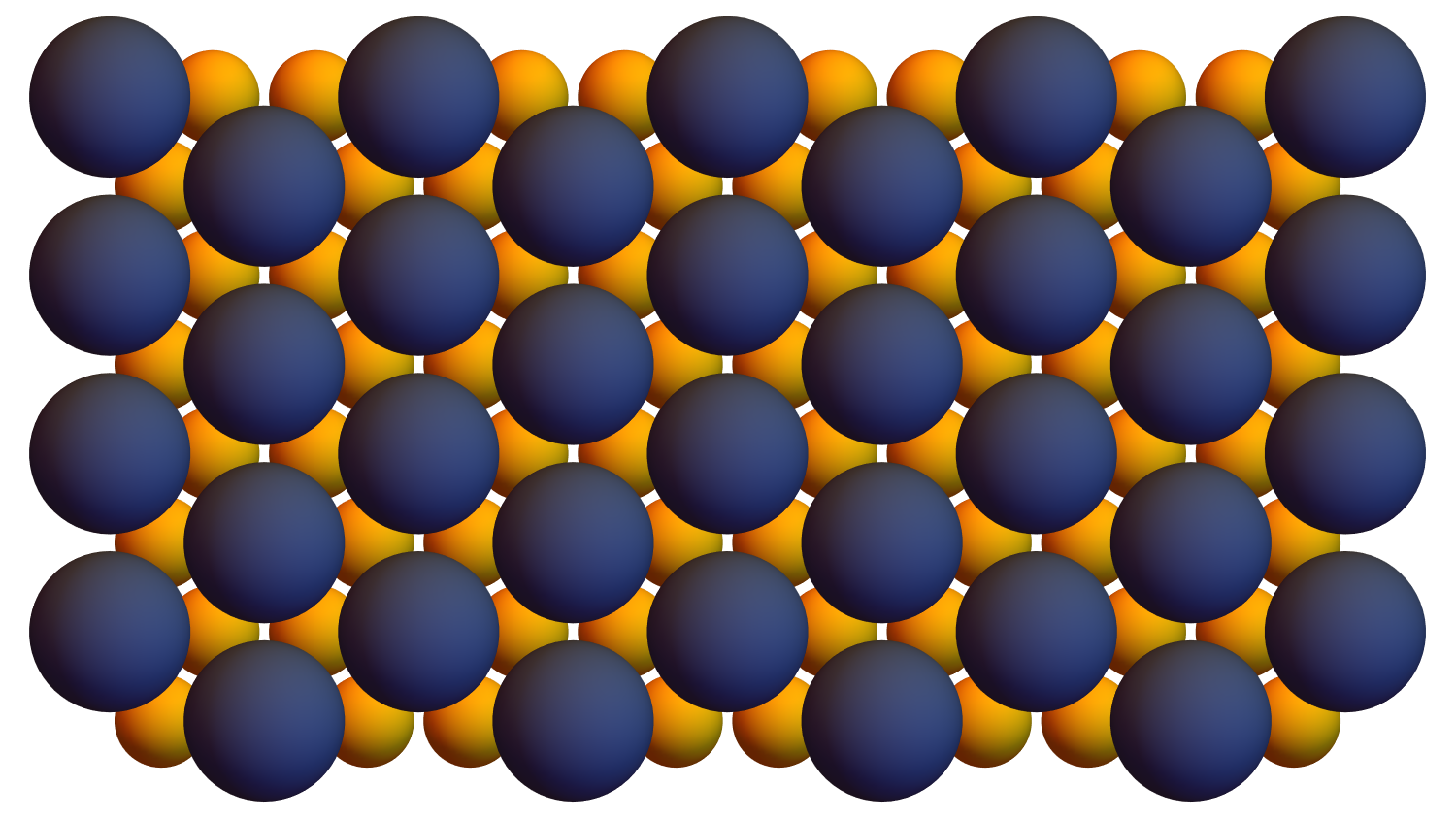}
    \end{minipage}
    \begin{minipage}[c]{0.49\linewidth}
    \centering
        \includegraphics[width=0.6\linewidth]{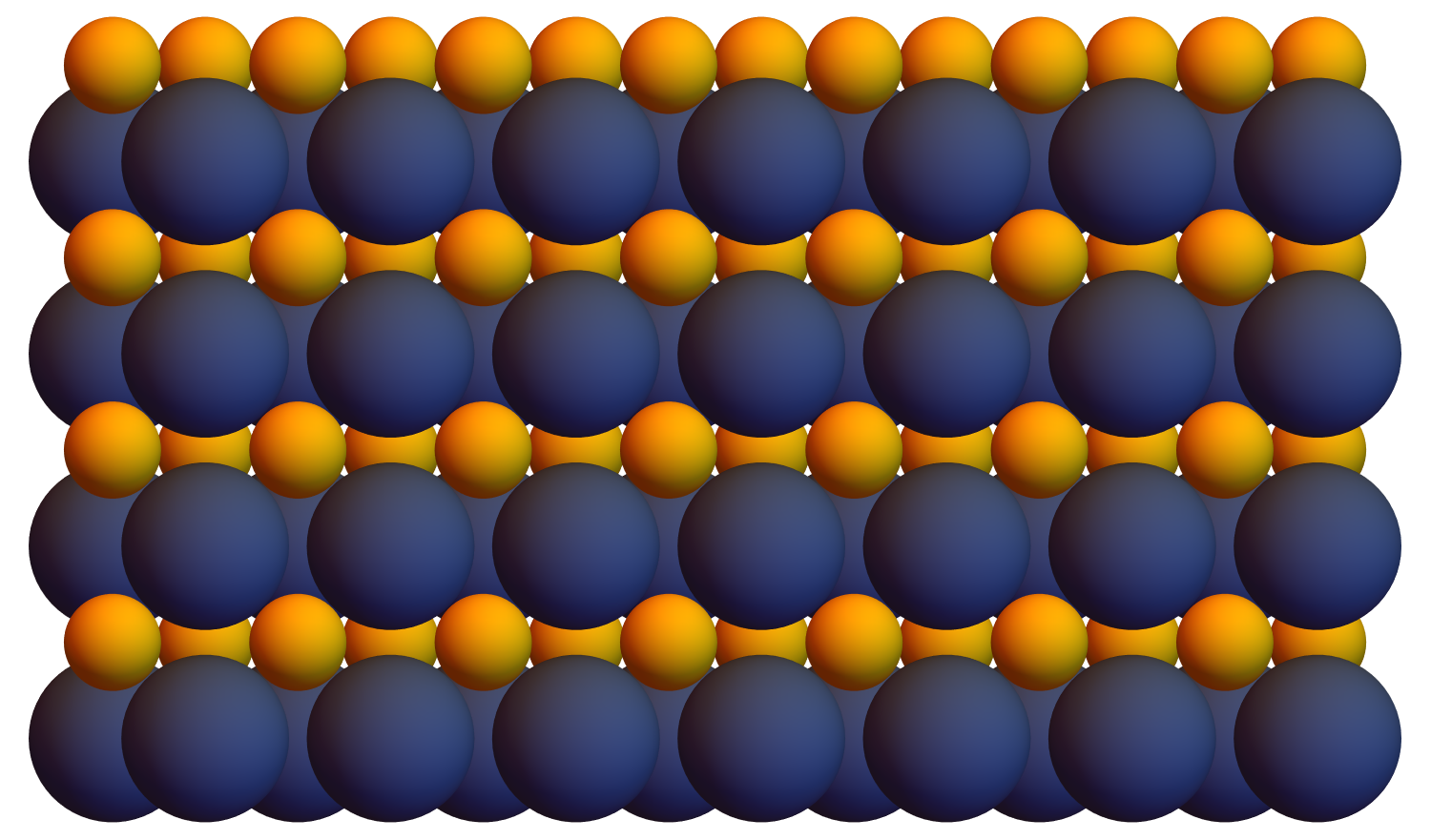}   
    \end{minipage}

 \begin{minipage}[t]{0.49\linewidth}
        \RaggedRight b)
    \end{minipage}
    \begin{minipage}[t]{0.49\linewidth}
        \RaggedRight e)
    \end{minipage}
    
     \begin{minipage}[c]{0.49\linewidth}
     \centering
     \includegraphics[width=0.9\linewidth]{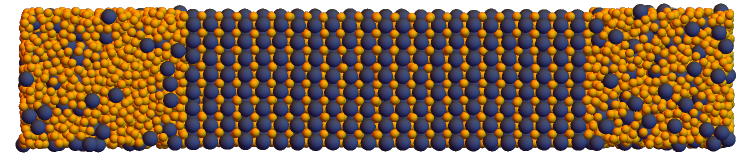}
    \end{minipage}
    \begin{minipage}[c]{0.49\linewidth}
    \centering
        \includegraphics[height=2.3cm]{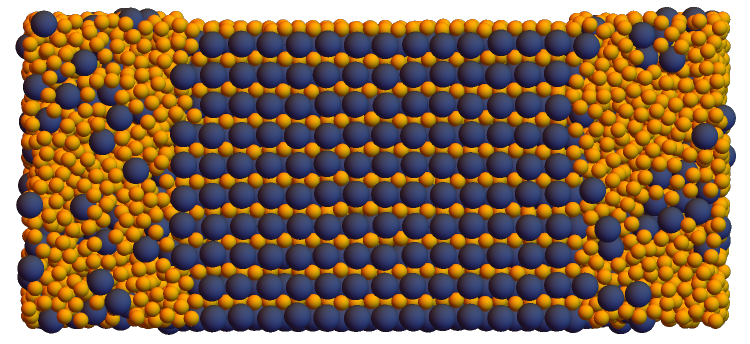}  
    \end{minipage}
  \begin{minipage}[t]{0.49\linewidth}
        \RaggedRight c)\\
        \centering
        \includegraphics[height=5.5cm]{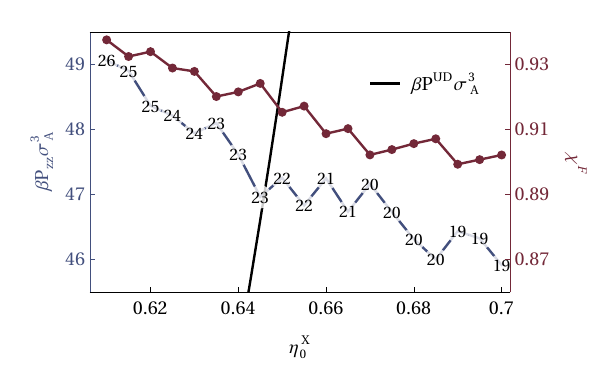}
    \end{minipage}
    \begin{minipage}[t]{0.49\linewidth}
        \RaggedRight f)\\
        \centering
        \includegraphics[height=5.5cm]{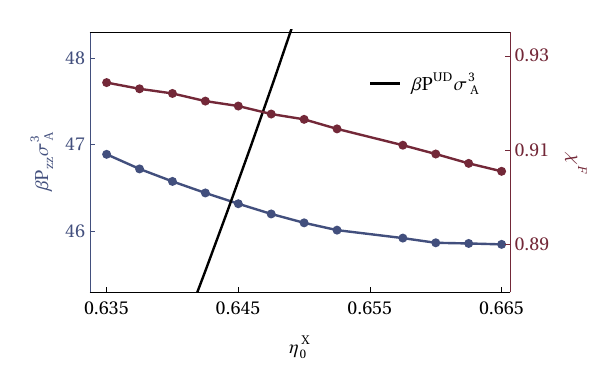}
    \end{minipage}

    \caption[width=1\linewidth]{The effect of different AB$_{2}$ crystal orientations on the direct-coexistence simulations. In the first orientation, the crystal is aligned such that the crystal plane shown in a) is parallel to the interface. A representative snapshot at $(\eta^G, \chi^G) = (0.585, 0.8)$ and $\eta^X_0 = 0.64$ is shown in b). For the same global parameters, c) shows the normal pressure component $P^G_{zz}$ and fluid composition $\chi^F$ as functions of the initial crystal packing fraction $\eta^X_0$. The numbers on the pressure curves indicate the number of crystal layers in the crystal phase, which remained constant for most of the simulation. The black line represents the bulk crystal pressure. In the second orientation, the crystal is rotated by $\pi/2$ relative to the first case, such that the crystal plane shown in d) is parallel to the interface. A representative snapshot at $(\eta^G, \chi^G) = (0.58, 0.8)$ and $\eta^X_0 = 0.64$ is shown in e), while the corresponding pressure and composition as functions of $\eta^X_0$ are shown in f). These results highlight the difference in interfacial stiffness between the two crystal orientations. Note that the two orientations are associated with slightly different global packing fractions, which leads to a systematic shift in pressure between the two systems.}
    \label{fig:sturdyinterfaceAB2} 
\end{figure*}

\begin{figure*} \begin{minipage}[t]{0.49\linewidth}
        \RaggedRight a)\\
        \centering
        \includegraphics[width=0.6\linewidth]{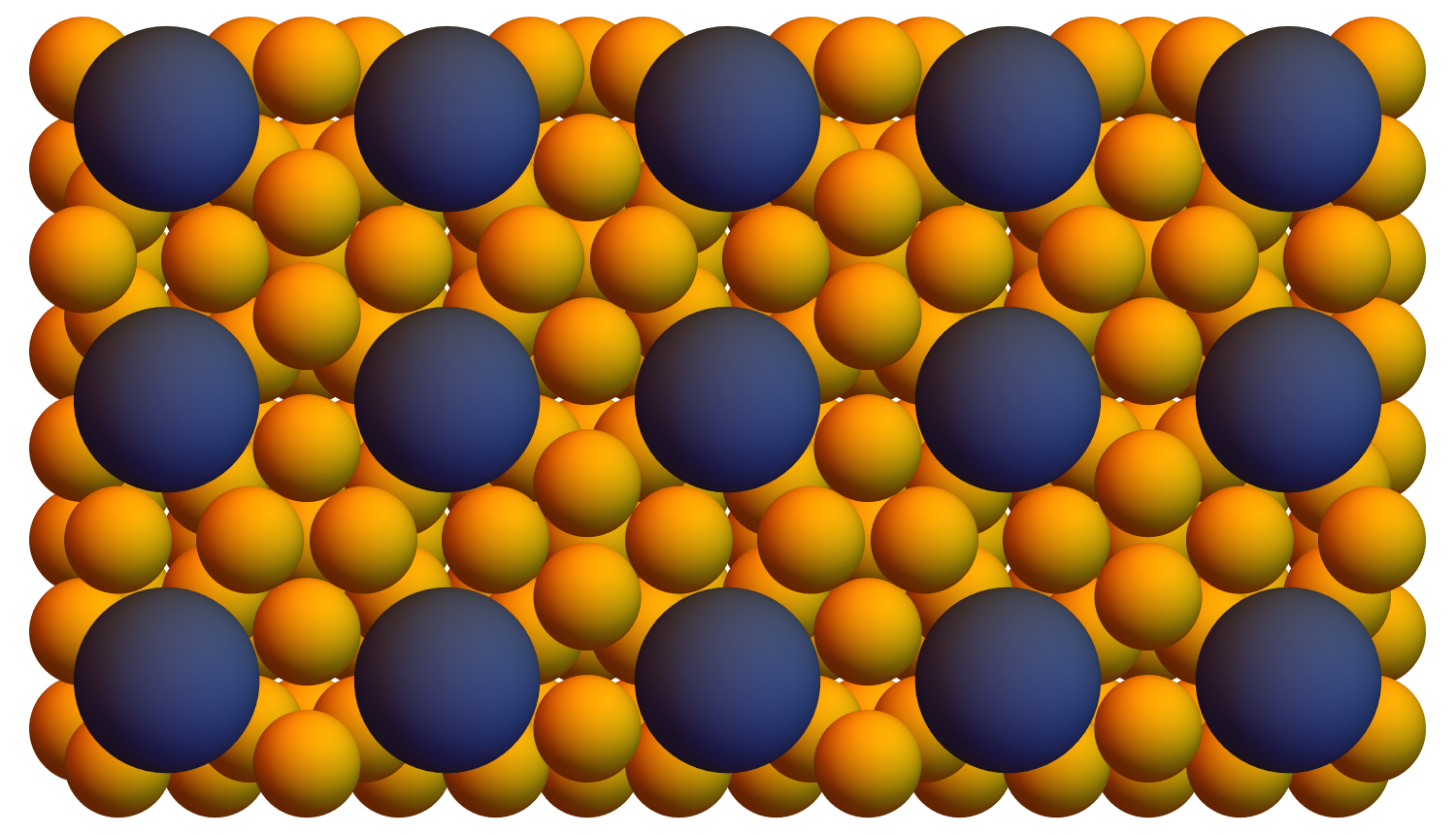}
    \end{minipage}
    \begin{minipage}[t]{0.49\linewidth}
        \RaggedRight d)\\
        \centering
        \includegraphics[width=0.6\linewidth]{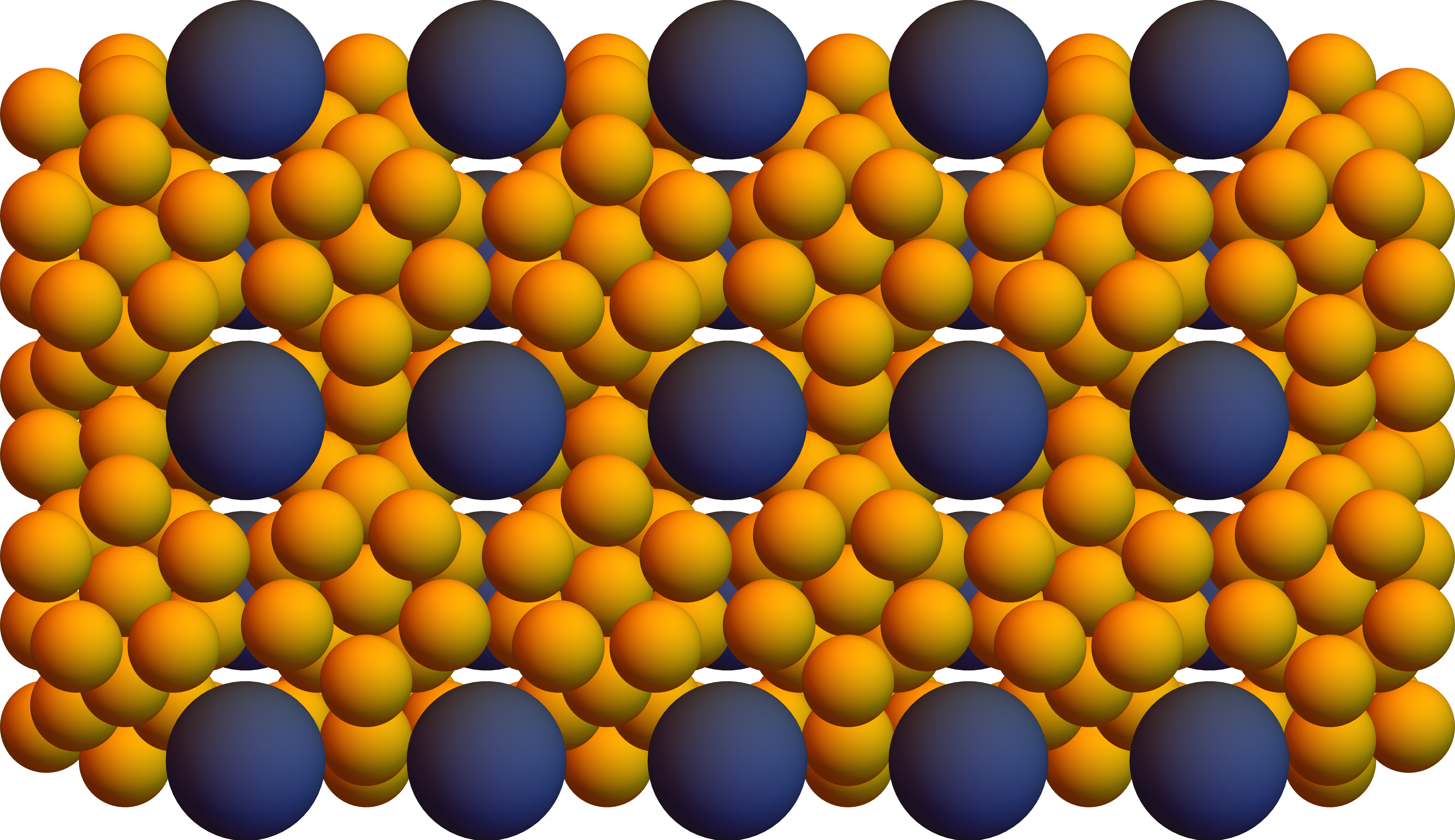}
    \end{minipage}
    
 \begin{minipage}[t]{0.49\linewidth}
        \RaggedRight b)
    \end{minipage}
    \begin{minipage}[t]{0.49\linewidth}
        \RaggedRight e)
    \end{minipage}
    
     \begin{minipage}[c]{0.49\linewidth}
     \includegraphics[width=0.9\linewidth]{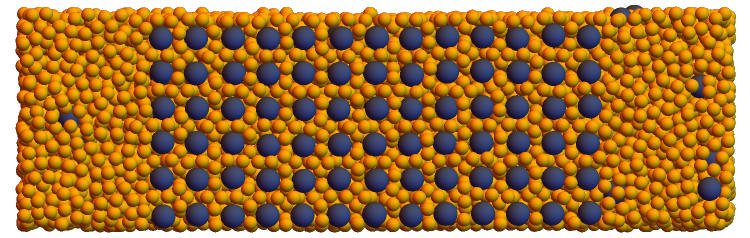}
    \end{minipage}
    \begin{minipage}[c]{0.49\linewidth}
        \includegraphics[height=2.4cm]{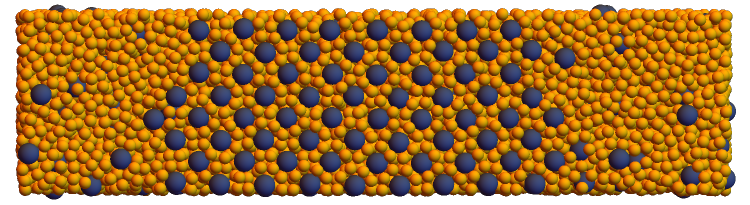}   
    \end{minipage}
  \begin{minipage}[t]{0.49\linewidth}
        \RaggedRight c)\\
        \centering
        \includegraphics[height=5.5cm]{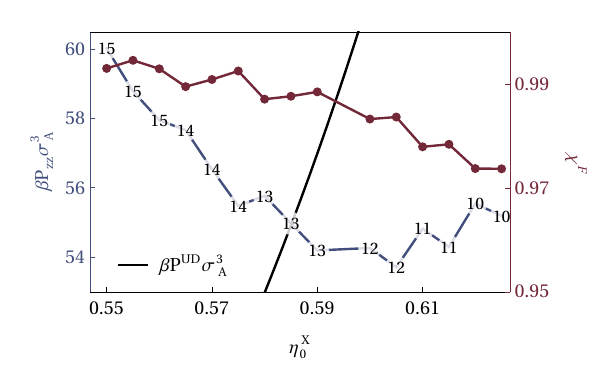}
    \end{minipage}
    \begin{minipage}[t]{0.49\linewidth}
        \RaggedRight f)\\
        \centering
        \includegraphics[height=5.5cm]{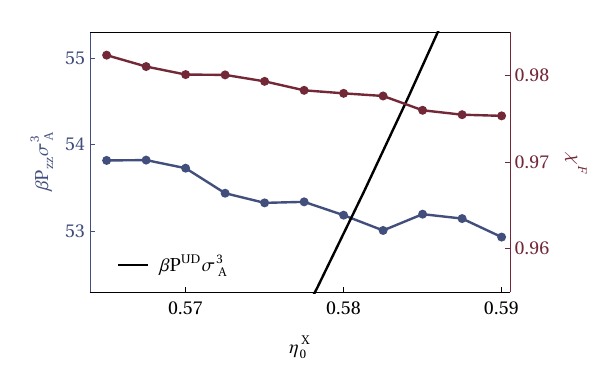}
    \end{minipage}
    
 
    \caption[width=1\linewidth]{The effect of different AB$_{13}$ crystal orientations on the direct-coexistence simulations. In the first orientation, the crystal is aligned such that the crystal plane shown in a) is parallel to the interface, with alternating particle layers perpendicular to the interface. A representative snapshot at $(\eta^G, \chi^G) = (0.55, 0.95)$ and $\eta^X_0 = 0.59$ is shown in b). For the same global parameters, c) shows the normal pressure component $P^G_{zz}$ and fluid composition $\chi^F$ as functions of the initial crystal packing fraction $\eta^X_0$. The numbers on the pressure curves indicate the number of crystal layers in the crystal phase, which remained constant for most of the simulation. The black line represents the bulk crystal pressure. In the second orientation, the crystal is rotated by $\pi/4$ relative to the first case, such that the crystal plane shown in d) is parallel to the interface. A representative snapshot at $(\eta^G, \chi^G) = (0.54, 0.95)$ and $\eta^X_0 = 0.59$ is shown in e), while the corresponding pressure and composition as functions of $\eta^X_0$ are shown in f). These results highlight the difference in interfacial stiffness between the two crystal orientations. Note that the two orientations are associated with slightly different global packing fractions, leading to a systematic shift in pressure between both systems.}
    \label{fig:sturdyinterfaceAB13} 
\end{figure*}

\begin{figure*}
        \includegraphics[width=\linewidth]{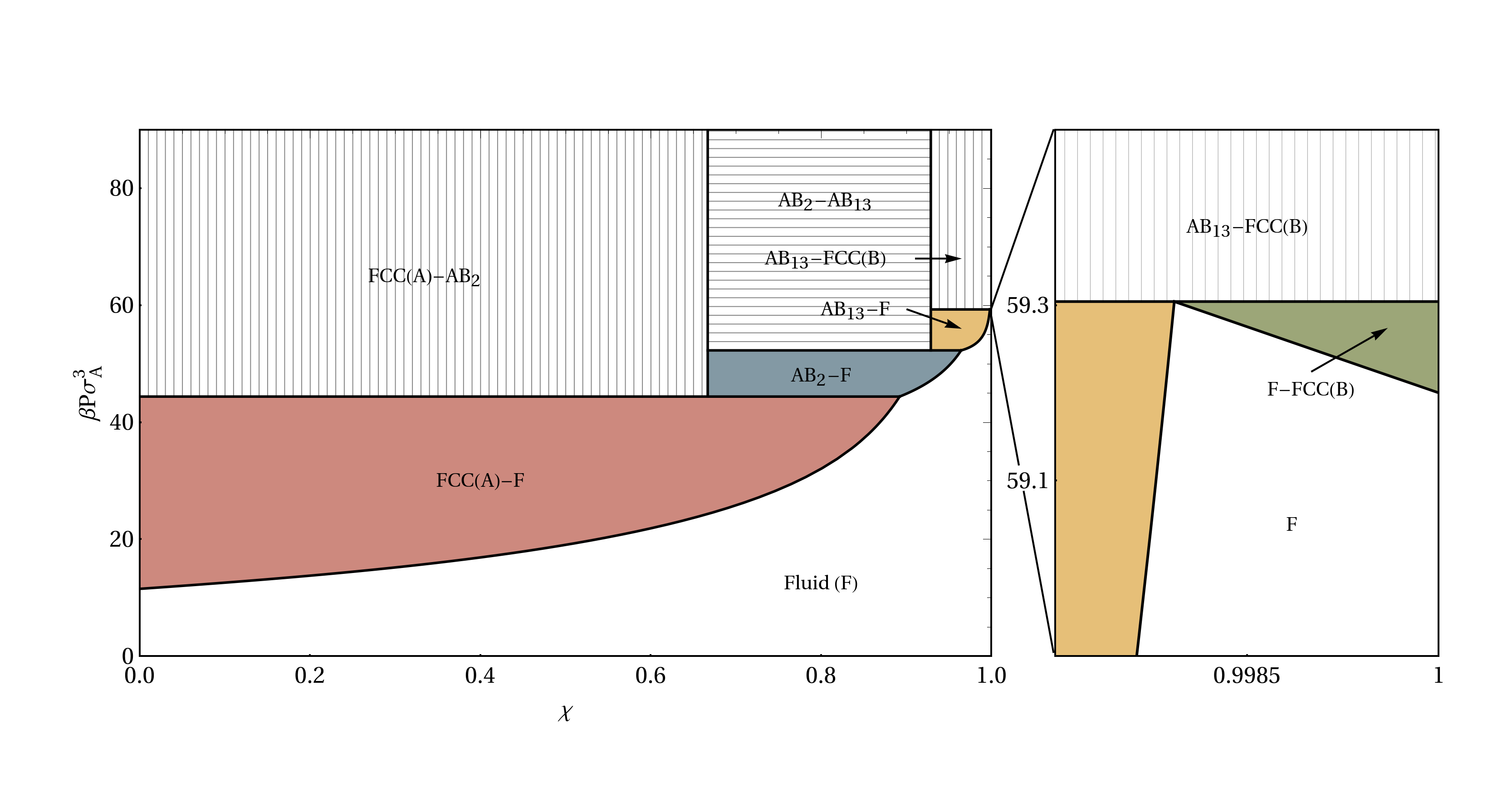}
    \caption[width=1\linewidth]{Phase diagram of a binary mixture with size ratio $q = 0.58$, obtained from direct-coexistence simulations. The right panel shows a zoomed-in view of the phase diagram, indicating that there is a small but finite region where the F-FCC(B) coexistence is stable. In Appendix ~\ref{app:fittedpolynomials}, we display the phase boundaries fits.}
    \label{fig:newphasediagram} 
\end{figure*}

Finally, we note that care must be taken when selecting the $\eta_0^X$ interval for fluid-FCC(A) coexistences. In our simulations, we observed that crystals that were deformed too much became distorted and developed many defects. While defects can occur in all deformed crystals, the effect was especially pronounced in the FCC(A) crystal. In Appendix~\ref{app:appdefects} we show an example of a planar defect that appeared at multiple state points under large deformation. These issues are avoided by only considering $\eta_0^X$ close to the coexistence packing fraction. Note that by visualizing the simulation, it can easily be checked by eye whether the crystal contains a significant number of defects.

\subsection{AB$_\mathbf{2}$}
To determine the fluid-AB$_{2}$ phase boundary we considered direct coexistences for global packing fractions in the range $\eta^G\in[0.56, 0.59]$ and global compositions in the range $\chi^G\in[0.75, 0.85]$, for systems initially consisting of $N_X = 3600$ crystal particles. In all simulations the number of crystal unit cells in $x$- and $y$-direction was set to 10. 
We found that the behavior of the fluid-crystal interface was highly sensitive to the choice of crystal plane in contact with the fluid. Using an orientation where alternating particles layers lie parallel to the interface, see Figs.~\ref{fig:sturdyinterfaceAB2}a) and \ref{fig:sturdyinterfaceAB2}b), resulted in a crystal that only grew or melted in discrete layers.

In Fig. \ref{fig:sturdyinterfaceAB2}c), we show the pressure and fluid composition at $\eta^G=0.585$ and $\chi^G = 0.8$ for a direct coexistence with this crystal orientation. In the figure, we also indicate the number of crystal layers present during each simulation. These figures reveal that the interface is indeed very stiff, with a very small interfacial width. Consequently, the simulation effectively only samples integer numbers of crystal layers, which strongly limits the system's ability to adapt the number of particles in the crystal phase as necessary to sample the equilibrium coexistence. This results in the stepwise behaviour of both $\beta P_{zz}\sigma_A^3$ and $\chi^F$ of Fig. \ref{fig:sturdyinterfaceAB2}c). 

As a way of addressing this issue, we found that rotating the crystal by $\pi/2$, such that the planes of alternating particles lie perpendicular to the interface (see Figs.~\ref{fig:sturdyinterfaceAB2}d) and \ref{fig:sturdyinterfaceAB2}e)), yielded a significantly less stiff interface. Fig. \ref{fig:sturdyinterfaceAB2}f) shows the pressure and fluid composition in a direct-coexistence simulation at $\eta^G=0.58$ and $\chi^G = 0.8$ with this crystal orientation. From the representative snapshot of this system, displayed in Fig. \ref{fig:sturdyinterfaceAB2}e), we can indeed observe that the interface is much wider. Hence, to calculate the fluid-AB$_2$ coexistence line, we use this orientation in all our simulations.
In Fig. \ref{fig:phasediagramperphase}c) we plot the resulting phase boundary. Again we see that, compared to Ref. \onlinecite{eldridge1993entropy}, the boundary is shifted downward consistently.

\subsection{AB$_\mathbf{13}$}
To determine the fluid-AB$_{13}$ phase boundary we considered direct coexistences for global packing fractions in the range $\eta^G\in[0.53, 0.55]$ and global compositions in the range $\chi^G\in[0.94, 0.9625]$, for systems initially consisting of $N_X = 5376$ crystal particles. 

Similar to the AB$_2$ crystal, we found that the results are very sensitive to the crystal plane that is in contact with the fluid. To illustrate this, we looked at two different orientations of the AB$_{13}$ crystal structures, shown in Figs.~\ref{fig:sturdyinterfaceAB13}a),b), and \ref{fig:sturdyinterfaceAB13}d), e). For the first orientation, Fig.~\ref{fig:sturdyinterfaceAB13}c) shows the pressure in the $z$-direction. Analogous to the fluid-AB$_2$ case, we observe crystal growth or melting to occur only through the addition or removal of entire layers. Rotating the crystal by $\pi/4$, led to a less stiff interface and a smoother behavior of the pressure, as shown in Fig.~\ref{fig:sturdyinterfaceAB13}f). Hence, we use this orientation in our phase boundary determinations. 
In Fig. \ref{fig:phasediagramperphase}d), we plot the resulting phase boundary, which again is consistently shifted downward compared to Ref. \onlinecite{eldridge1993entropy}.

Finally, we note that finite-size effects associated with the amount of fluid in the simulation box play a noticeable role in the fluid-AB$_{13}$ direct-coexistence simulations. We observed that the addition of a single crystal unit cell occurs very slowly, particularly in systems containing relatively few $A$ particles. Because the AB$_{13}$ unit cell is large, the insertion or removal of even one cell causes substantial changes in both the fluid density and composition. As a result, the simulations exhibit slow and pronounced fluctuations in pressure and composition. Consequently, long simulation times are required to sample the system sufficiently. Note that these difficulties are not specific to our implementation, but rather are inherent to direct-coexistence methods in general. Although we expect that increasing the system size or extending the simulation time would likely further improve the precision of the extracted phase boundaries, we consider the results presented here to be sufficiently accurate and reliable.

\subsection{Phase diagram}
In Fig. \ref{fig:newphasediagram}, we show the phase diagram obtained using the direct-coexistence method. The zoomed-in inset of the same diagram highlights the small but finite region where fluid–FCC(B) coexistence is stable. In Appendix ~\ref{app:fittedpolynomials}, the functions that fit the phase boundaries are given.

\section{Conclusions}
In conclusion, we have extended the direct-coexistence method introduced by Ref.~\onlinecite{smallenburg2024simple} to determine phase boundaries in binary mixtures that form stoichiometric crystals. It is worth noting that, in addition to the $NVT$  direct-coexistence simulations used here, binary mixtures would also yield stable coexistences in the $NP_zT$ ensemble\cite{noya2008determination,espinosa2013fluid,zykova2010monte}. In this ensemble, the pressure in the $z$-direction, $P_z$, is fixed to the value consistent with the lattice spacing imposed by the box dimensions in $x$ and $y$. In principle, this ensemble has the advantage that only one simulation is required to identify the strain-free crystal. However, in practice, the requirement of multiple simulations is not particularly expensive, since they can easily be performed simultaneously. Moreover, since constant-pressure simulations often have long correlation times in the volume fluctuations, it is likely that the simulation in the $NP_zT$ -ensemble would take significantly longer than a single simulation in the $NVT$-ensemble.

The direct-coexistence method discussed in this paper offers several clear advantages over traditional free-energy based approaches. As in the monodisperse case, the method is less prone to numerical error since it does not require integrating over a series of measurements. In addition, the method is straightforward to implement and to verify. A key advantage specific to binary hard-sphere mixtures, is that it does not require knowledge of the EOS of state of the binary fluid. Free-energy calculations either require measuring the EOS over a wide range of compositions and densities, or require relying on some semi-empirical expressions like the BMCSL EOS. As integration over the EOS is required, small errors in the EOS can lead to significant shifts in the predicted phase boundaries as we have shown in this paper.

There are certain aspects that require caution when performing direct-coexistence simulations. In general, direct coexistence simulations will be computationally more expensive than free-energy calculations, due to the slow relaxation of the explicit interfaces.  
Specifically for this system, we observed that both the AB$_{2}$ and AB$_{13}$ crystals are sensitive to the orientation of the crystal in the box: some orientations resulted in extremely stiff interfaces, where growth or melting only occurred in discrete layers. In this work, we identified crystal orientations that led to less stiff interfaces. 
We speculate that for most crystals, the best choice of crystal plane in contact with the fluid is a plane that does not show strong layering in the direction normal to the interface. This may run counter to the natural inclination to use a ``clean'' crystal plane as the interface, but --- as our result show --- could significantly impact the efficiency and accuracy of sampling in direct-coexistence simulations.

In short, we conclude that the direct-coexistence method presented here and in Ref.~\onlinecite{smallenburg2024simple} provides a robust approach that is simple to implement and that leads to an efficient and accurate determination of phase boundaries in the class stoichiometric binary crystals.

\section*{Acknowledgments}
L.F. acknowledges funding from the Dutch Research Council (NWO) as part of the Vici ENW  program with file number  VI.C.242.116 and grant ID \href{https://doi.org/10.61686/PWYNK61282}{https://doi.org/10.61686/PWYNK61282}. F.S. acknowledges funding from the  Agence Nationale de la Recherche (ANR), grant ANR-21-CE30-0051. 

\section*{Data Availability Statement}
All simulation codes needed to reproduce the data, as well as all relevant simulation data and notebooks to analyze the data and generate the figures are published as a data package in Ref.~\onlinecite{alkemade_2026_18681499}.

\section{References}
\addcontentsline{toc}{part}{Bibliography}
\markboth{\MakeUppercase{Bibliography}}{}
\bibliographystyle{abbrvunsrt2}
\bibliography{refs}

\appendix
\section{Ratio lattice constants AB$_\mathbf{2}
$}
\label{app:latticeratio}
\begin{figure}
        \includegraphics[width=\linewidth]{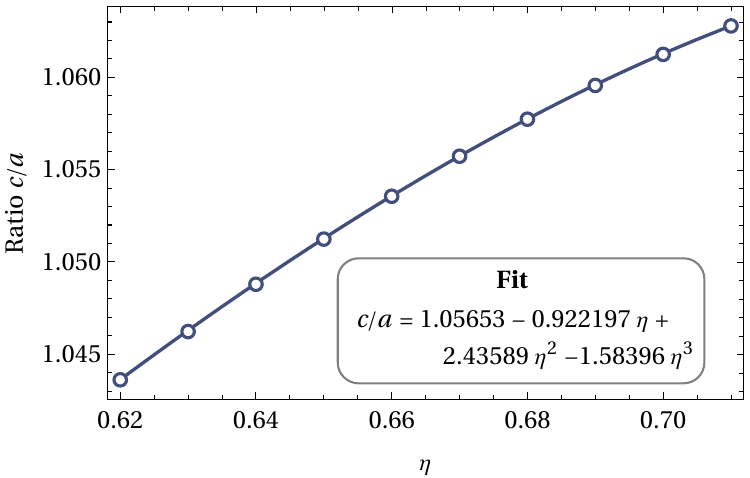}
    \caption[width=1\linewidth]{Ratio between the two defining lattice parameters $a$ and $c$ of the AB$_2$ lattice (see Fig.\ref{fig:crystals}) associated with an isotropic crystal pressure, as a function of the packing fraction. The fitted solid line is used in all AB$_2$ simulations. Measurements are based on simulations of an AB$_2$ crystal consisting of $N=1536$ particles. Simulations are run for $t/\tau = 10^5$.}
    \label{fig:lattice ratio} 
\end{figure}

The unit cell of an AB$_2$ crystal is defined by two parameters $a$ and $c$, see Fig.~\ref{fig:crystals}, whose ratio will depend on the density.
To find the equilibrium ratio $c/a$ for which the AB$_2$ crystal is strain-free, we performed a series of simulations with different $c/a$-ratio's and identified where the pressure was isotropic. In Fig. \ref{fig:lattice ratio}, we plot these equilibrium $c/a$-values as a function of packing fraction. In addition, the figure displays the $c/a$-fit that we use throughout all AB$_2$ simulations. 


\begin{figure*}[t!]\centering
 \begin{minipage}[t]{0.4\linewidth}
        \RaggedRight a)\\
        \includegraphics[width=\linewidth]{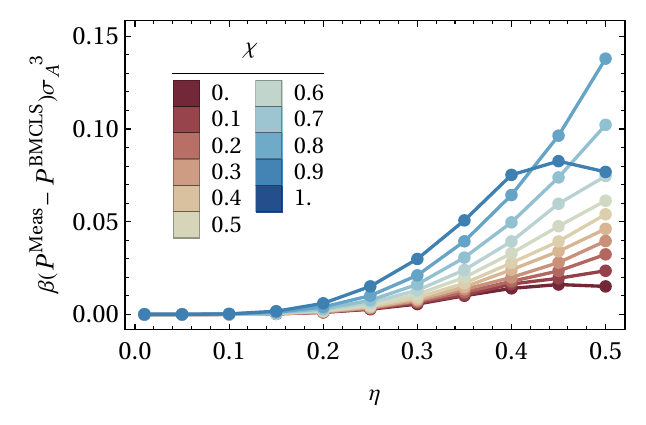}
    \end{minipage}
    \begin{minipage}[t]{0.4\linewidth}
        \RaggedRight b)\\
        \includegraphics[width=\linewidth]{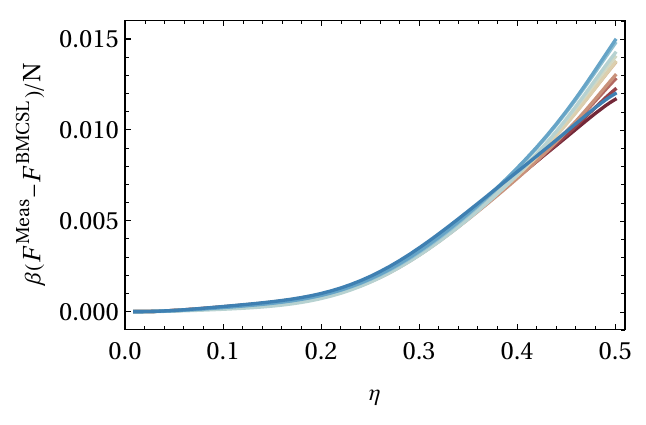}
    \end{minipage}
    \caption[width=1\linewidth]{a) Difference between the equation of state (EOS) as measured in EDMD simulations and the BMCSL EOS \cite{boublik1970hard,mansoori1971equilibrium}. b) Difference in Helmholtz free energy using both EOS to calculate the free energy. All simulations consisted of $N =4000$ particles and ran for $10^5 \tau$. The  legends in panel a) also apply to panel b).}
    \label{fig:difPandF} 
\end{figure*}

\section{Accuracy BMCSL EOS}
\label{app:discrepancyBMCSL}
Since the BMCSL EOS \cite{boublik1970hard,mansoori1971equilibrium} is a semi-empirical expression, it is interesting to examine how well it agrees with data obtained in simulations. For a range of compositions, we therefore measured the EOS in EDMD simulations. From this EOS, we then obtained the free energy, where we fitted the measured pressures using a polynomial fit that explicitly incorporated the exact first and second virial coefficients for a binary mixture. In Fig. \ref{fig:difPandF}a) and b) we plot the difference between the EOS and the BMCSL EOS and the corresponding Helmholtz free energies, respectively. From the plot it is clear that using the BMCSL EOS leads to a small but significant deviation from the simulations in both the pressure and the free energy. As mentioned in the main paper, this underestimation of the free energy leads to an overestimation of the coexistence pressure and thus an upward shift in the phase boundaries. 

\section{Fitted functions phase boundaries}
\label{app:fittedpolynomials}
Below we provide fitted equations for the coexistence pressure as a function of the fluid composition 
$\chi^F$ for the four phases. These fits correspond to the solid lines displayed in Fig.~\ref{fig:phasediagramperphase}.

\begin{widetext}
\begin{align*}
&\beta P_{\text{FCC(A)}}\sigma_A^3 = \frac{11.5425 - 2.4829\chi^F- 6.9330(\chi^F)^2 + 6.5686(\chi^F)^3 - 7.2276(\chi^F)^4}{(1-\chi^F)^{1.0933}}\\
&\beta P_{\text{AB}_2}\sigma_A^3= \frac{-21.7760+158.3203\chi^F -408.5582(\chi^F)^2 +457.3318(\chi^F)^3 -185.3343(\chi^F)^4}{(\chi^F)^{5.8557}(1-\chi^F)^{1.1864}}\\
&\beta P_{\text{AB}_{13}}\sigma_A^3 = \frac{521.7075 -568.7023\chi^F-990.7700(\chi^F)^2+1376.1094(\chi^F)^3 -295.8303(\chi^F)^4}{(\chi^F)^{10.2500}(1-\chi^F)^{0.0535}}\\
&\beta P_{\text{FCC(B)}}\sigma_A^3 = \frac{6.8404 + 20.7608\chi^F+ 15.4567(\chi^F)^2 + 10.4413(\chi^F)^3 + 5.7010(\chi^F)^4}{(\chi^F)^{2.6332}}
\end{align*}
\end{widetext}

\section{Defects}
\label{app:appdefects}
\begin{figure}[h]
 \begin{minipage}[t]{\linewidth}
        \includegraphics[width=0.95\linewidth]{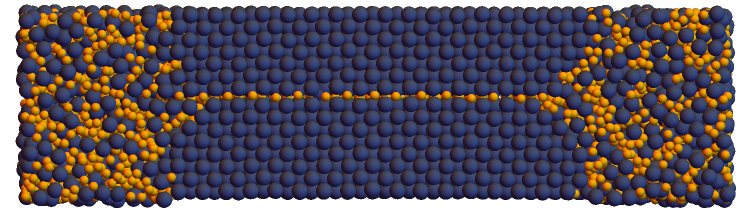}
    \end{minipage}
  \begin{minipage}{\linewidth}
        \includegraphics[width=0.95\linewidth]{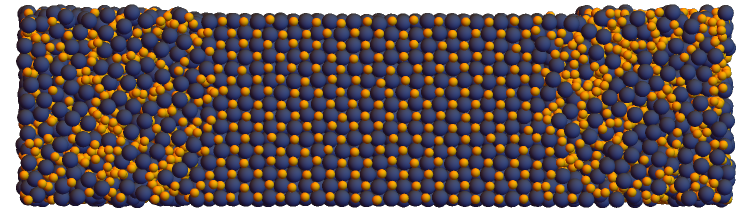}
    \end{minipage}\\

    \caption[width=1\linewidth]{Planar defect observed in a fluid-FCC(A) phase coexistence at a global packing fraction $\eta^G = 0.58$, global composition $\chi^G = 0.4$ and initial crystal packing fraction of $\eta_0^X= 0.69$. The upper snapshot displays a top view of the system, while the lower snapshot shows a cut-through of the system at the position of the defect.}
    \label{fig:FCCplanerdefect} 
\end{figure}

When the initial crystal density in a direct-coexistence simulation is far removed from the equilibrium density at coexistence, the most likely outcomes are full melting or full crystallization of the system. However, in some cases, an alternative response is the formation of large-scale defects due to the strain imposed on the crystal. In our case, this was most common in the FCC(A) phase. To illustrate the way these defects can manifest, in Fig. \ref{fig:FCCplanerdefect} we show a planar defect that was observed in the fluid-FCC(A) phase coexistence, where an entire layer of small particles invaded the crystal.

\end{document}